\documentclass[utf8]{FrontiersinHarvard}
\usepackage{url,hyperref,lineno,microtype,subcaption}
\usepackage[onehalfspacing]{setspace}
\usepackage[]{xspace}
\usepackage{amsmath}

\newcommand{\arcsec}{\ensuremath{^{\prime\prime}}}
\newcommand{\arcmin}{\ensuremath{^{\prime}}}

\newcommand{\chandra}{\textit{Chandra}\xspace}

\newcommand{\hexp}{\textit{\mbox{HEX-P}}\xspace}

\newcommand{\xmm}{\textit{XMM-Newton}\xspace}

\newcommand{\nustar}{\textit{NuSTAR}\xspace}
\newcommand{\athena}{\textit{Athena}\xspace}

\newcommand{\fluxcgs}{\ensuremath{{\rm erg~s}^{-1}~{\rm cm}^{-2}}}
\newcommand{\lumcgs}{\ensuremath{{\rm erg~s}^{-1}}\xspace}

\newcommand{\Mtwo}{M82\,X-2\xspace}
\newcommand{\fref}{Fig.~\ref}

\def\keyFont{\fontsize{8}{11}\helveticabold }
\def\firstAuthorLast{Bachetti {et~al.}}
\def\Authors{Matteo Bachetti\,$^{1,*}$,
Matthew J. Middleton\,$^{2}$,
Ciro Pinto\,$^{3}$,
Andr\'es G\'urpide\,$^{2}$,
Dominic J. Walton\,$^{4}$,
Murray Brightman\,$^{5}$,
Bret Lehmer\,$^{6}$,
Timothy P. Roberts$^{7}$,
Georgios Vasilopoulos$^{8}$,
Jason Alford\,$^{9}$,
Roberta Amato\,$^{10}$,
Elena Ambrosi\,$^{3}$,
Lixin Dai$^{11}$,
Hannah P. Earnshaw$^{5}$,
Hamza El Byad\,$^{1}$
Javier A. Garc\'ia$^{12,5}$,
Gian Luca Israel$^{13}$,
Amruta Jaodand$^{5}$,
Kristin Madsen$^{12}$,
Chandreyee Maitra$^{14}$,
Shifra Mandel\,$^{9}$,
Kaya Mori\,$^{9}$,
Fabio Pintore\,$^{3}$,
Ken Ohsuga\,$^{15}$,
Maura Pilia\,$^{1}$,
Daniel Stern$^{16}$,
George Younes$^{12}$,
Anna Wolter$^{17}$}

\def\Address{$^{1}$INAF-Osservatorio Astronomico di Cagliari, 
Selargius (CA), Italy \\
$^{2}$School of Physics and Astronomy, University of Southampton
UK\\
$^{3}$INAF – IASF Palermo, 
Italy \\
$^{4}$Centre for Astrophysics Research, University of Hertfordshire, 
UK \\
$^{5}$Cahill Center for Astrophysics, California Institute of Technology, 
USA \\
$^{6}$Department of Physics, University of Arkansas, 
USA \\
$^{7}$Centre for Extragalactic Astronomy \& Department of Physics, Durham University, 
UK\\
$^{8}$Department of Physics, National and Kapodistrian University of Athens, 
Greece \\
$^{9}$Columbia Astrophysics Laboratory, Columbia University, 
USA \\
$^{10}$IRAP, CNRS, Université de Toulouse, CNES, 
France \\
$^{11}$Department of Physics, University of Hong Kong, 
China \\
$^{12}$X-ray Astrophysics Laboratory, NASA Goddard Space Flight Center, 
USA \\
$^{13}$INAF – Osservatorio Astronomico di Roma, 
Italy\\
$^{14}$Max-Planck-Institut f\"ur extraterrestrische Physik, Gei{\ss}enbachstra{\ss}e 1, D-85748 Garching, Germany \\
$^{15}$Center for Computational Sciences, University of Tsukuba, 
Japan \\
$^{16}$ Jet Propulsion Laboratory, California Institute of Technology, 
USA  \\
$^{17}$INAF-Osservatorio Astronomico di Brera, 
Milano, Italy
}
\def\corrAuthor{Matteo Bachetti}

\def\corrEmail{matteo.bachetti@inaf.it}

\begin{document}
\onecolumn
\firstpage{1}

\title[ULXs with HEX-P]{The High Energy X-ray Probe (HEX-P): Studying Extreme Accretion with Ultraluminous X-ray Sources}

\author[\firstAuthorLast ]{\Authors}
\address{}
\correspondance{}

\extraAuth{}

\maketitle

\begin{abstract}

\section{}
\textbf{Introduction.} Ultraluminous X-ray sources (ULXs) represent an extreme class of accreting compact objects: from the identification of some of the accretors as neutron stars to the detection of powerful winds travelling at 0.1–0.2 c, the
increasing evidence points towards ULXs harbouring stellar-mass compact objects undergoing highly super-Eddington accretion. Measuring their intrinsic properties, such as the accretion rate onto the compact object, the outflow
rate, the masses of accretor/companion–hence their progenitors, lifetimes, and future evolution–is challenging due to ULXs being mostly extragalactic and in crowded fields. Yet ULXs represent our best opportunity to understand super-
Eddington accretion physics and the paths through binary evolution to eventual double compact object binaries and gravitational-wave sources.\\
\textbf{Methods.} Through a combination of end-to-end and single-source simulations, we investigate the ability of HEX-P to study ULXs in the context of their host galaxies and compare it to XMM-Newton and NuSTAR, the current instruments with the most similar capabilities.\\
\textbf{Results.} HEX-P's higher sensitivity, which is driven by its narrow point-spread function and low background, allows it to detect pulsations and broad spectral features from ULXs better than XMM-Newton and NuSTAR.\\
\textbf{Discussion.} We describe the value of HEX-P in understanding ULXs and their associated key physics, through a
combination of broadband sensitivity, timing resolution, and angular resolution, which make the mission ideal for pulsation detection and low-background, broadband spectral studies.

\tiny
 \keyFont{ \section{Keywords:} ultraluminous X-ray sources, HEX-P, pulsars, black holes, accretion, spectra} 
\end{abstract}

\twocolumn

\section{Introduction}

Ultraluminous X-ray sources (ULXs, see \citealt{kaaretUltraluminousXRaySources2017,fabrikaUltraluminousXRaySources2021,kingUltraluminousXraySources2023,PintoWalton2023ULXwindsrev} for recent reviews) are off-nuclear X-ray sources whose apparent luminosities exceed the Eddington limit for a stellar-mass black hole (e.g., $\sim10^{39}$\,\lumcgs for a $\sim$10$M_{\odot}$ black hole).

ULXs commonly have X-ray spectra consisting of two thermal components; the lower-energy component has a characteristic temperature of $<1$ keV, whereas the higher-energy component shows a cut off at $>3$ keV.
The low-energy component was initially interpreted as evidence for standard, sub-Eddington accretion onto intermediate-mass black holes \citep{millerXRaySpectroscopicEvidence2003}.
However, the cutoff, hinted at in high-quality observations with \xmm since the mid-2010s \citep{Roberts2005, Stobbart2006,2015aMiddleton}, was interpreted as Comptonization of disk photons by an optically thick corona and super Eddington accretion \citep[e.g.,][]{Gladstone2009}.
Hard X-ray coverage provided by \nustar \citep{Harrison2013} gave the highest-significance detections of this higher-energy cutoff, confirming the likely super-Eddington nature of most ULXs \citep[e.g.,][]{ bachettiUltraluminousXRaySources2013,waltonExtremelyLuminousVariable2013,ranaBroadbandXMMNewtonNuSTAR2015}.
Following the first detection of pulsations in M82 X-2 with \nustar (\citealt{bachettiUltraluminousXraySource2014}), a growing number of pulsating ULXs (PULXs, also referred to as ultraluminous pulsars, ULPs; e.g. \citealt{furstDiscoveryCoherentPulsations2016,israelAccretingPulsarExtreme2017,carpanoDiscoveryPulsationsNGC2018,rodriguezcastilloDiscoveryPulsarDay2020}) have been found, showing that neutron stars (NSs) are able to radiate at hundreds of times their Eddington limit.
The pulsations tend to be associated with an additional hard spectral component above 10 keV, which is reminiscent of the hard, curved spectra found in accreting pulsars \citep[e.g.,][]{pintorePulsatorlikeSpectraUltraluminous2017,waltonEvidencePulsarlikeEmission2018}.

The detection of some Galactic super-Eddington NSs provide a possible link between high-luminosity X-ray binaries (XRBs) and ULXs \citep[][e.g. Swift J0243+6124]{wilson-hodgeNICERFermiGBM2018}; in general, studying the highest-luminosity end of Galactic XRBs in the Milky Way Galaxy and external galaxies might provide clues on the onset of the ULX regime (see \citealt{fabbianoPopulationsXRaySources2006a} for a review, and \citet{connorsHighEnergyXray2023}, for \hexp's contribution to studies of this Galactic population).

Except for the few cases where pulsations have been observed, implying a neutron star (NS) accretor, determining the nature of the compact object in ULXs remains a major challenge for current observatories and a key open question for the vast majority of sources.
One way to address this issue is through population synthesis studies, which suggest that NS-ULXs may dominate the ULX population (\citealt{Middleton_2017_demographics_from_beaming, kingPulsingULXsTip2017,Wiktorowicz1904,Khan2021}).
PULXs can reach luminosities above $10^{41}$\,\lumcgs (e.g.
NGC~5907 X-1, \citealt{israelAccretingPulsarExtreme2017}), and given the similarity of PULX spectra to much of the ULX population, NSs might indeed power the majority of ULXs (\citealt{pintorePulsatorlikeSpectraUltraluminous2017, waltonEvidencePulsarlikeEmission2018}). In addition, a very small number of ULXs have shown  evidence for cyclotron resonance scattering features (CRSFs) in their spectra\footnote{See \citet{ludlamHighEnergyXray2023}, for CRSF studies on bright Galactic sources with \hexp} (e.g. \citealt{waltonPotentialCyclotronResonant2018,brightmanMagneticFieldStrength2018}), providing another route to identifying the presence of a NS, as well as a direct estimate of the magnetic field strength (likely indicating a weaker dipole but stronger multipolar field) close to the NS surface (\citealt{Middleton2019M51,  Kong2022_CRSF}).

The mechanism for the very high luminosities of ULXs is still debated. It is likely that the emission is partially collimated by an optically thick, radiatively driven outflow \citep{kingUltraluminousXRaySources2001} launched from the large scale-height disk expected at very high mass-transfer rates \citep{shakuraBlackHolesBinary1973}.
Indeed, many ULXs show evidence of high-velocity winds and outflows likely inflating the $\sim100$\,pc interstellar bubbles found around many ULXs at various wavelengths (\citealt{Stobbart2006,2014Middleton, 2015bMiddleton, pintoResolvedAtomicLines2016,Pakull2002,Gurpide2022,belfioreDiffuseXrayEmission2020}, see Section~\ref{sec:mass}).
However, these mass-loaded winds likely affect our ability to locate pulsing ULXs \citep[e.g.][]{mushtukovOpticallyThickEnvelopes2017}, and collimation alone is difficult to reconcile with the high pulsed fraction concurrent with the high luminosities of PULXs \citep[e.g.][]{israelAccretingPulsarExtreme2017}.
Another possibility is that a locally strong, possibly non-dipolar, magnetic field is capable of altering the local Thomson cross section \citep{baskoLimitingLuminosityAccreting1976,dallossoNuSTARJ09555169402015,eksiUltraluminousXraySource2015, mushtukovMaximumAccretionLuminosity2015,israelAccretingPulsarExtreme2017,briceSupereddingtonEmissionAccreting2021}.
These processes might also operate together to a greater or lesser extent.

Most ULXs are known to have other bright X-ray sources within 1\arcmin, in particular those located outside the Local Group (see Section~\ref{sec:population}). A clear detection and isolation of sources is therefore not possible given
the low resolution power of the high energy detectors in use, impeding a profound and detailed study of the whole population of ULXs.
The understanding of ULXs could reach considerable advancement by the combination, in one instrument, of high angular resolution, high time resolution and sensitivity in the hard X-rays.

The {\it High-Energy X-ray Probe} (\hexp; Madsen et al., in prep.) is a probe-class mission concept that offers sensitive broad-band X-ray coverage ($0.2-80$\,keV) with an exceptional combination of spectral, timing and angular resolution capabilities. It features two high-energy telescopes (HETs) that focus hard X-rays, and one low-energy telescope (LET) that focuses lower energy X-rays.

The LET consists of a segmented mirror assembly, coated with iridium on monocrystalline silicon that achieves a half power diameter (HPD) of 3.5\arcsec, and a low-energy DEPFET detector, of the same type as the Wide Field Imager \citep[WFI;][]{Meidinger2020} onboard \athena \citep{Nandra2013}. It has 512 $\times$ 512 pixels that cover a field of view of 11.3\arcmin $\times$ 11.3\arcmin, an effective passband of $0.2-25$\,keV, and a full frame readout time of 2\,ms, which can be operated in a 128 and 64 channel window mode for higher count-rates, to mitigate pile-up and allow for faster readout. Pile-up effects remain below an acceptable limit of $\sim 1\%$ for fluxes up to $\sim 100$\,mCrab -- a lot higher than typical extragalactic ULX fluxes -- in the smallest window configuration. The high angular resolution and low background result in a factor $\sim$2 sensitivity improvement with respect to \xmm.

The HET consists of two co-aligned telescopes and detector modules. The optics are made of Ni-electroformed full shell, mirror substrates, leveraging the heritage of \xmm \citep{Jansen2001}, and coated with Pt/C and W/Si multilayers for an effective passband of $2-80$\,keV. The high-energy detectors are of the same type flown on \nustar \citep{Harrison2013}, consisting of 16 CZT sensors per focal plane, tiled 4$\times$4, for a total of 128$\times$128 pixels spanning a field of view of 13.4\arcmin$\times$13.4\arcmin. This improvement in angular resolution allows for a much more sensitive instrument configuration compared to \nustar.

The broad \hexp X-ray passband and superior sensitivity compared with existing facilities provides a powerful and important opportunity to study ULXs across a wide range of energies, luminosities, and timescales.

\section{Challenges and open questions}

A number of major questions in the field of ULXs and super-Eddington accretion remain unanswered, preventing us from achieving a comprehensive understanding of their phenomenology and detailed accretion physics. We briefly describe them here.

\subsection{Where does the transferred mass go?} \label{sec:mass}
Understanding the fraction of matter being expelled in winds versus the fraction accreted is key to estimate the possible evolutionary paths of massive binary systems -- for instance, the formation of double compact object binaries and systems that create large ionising flux at high redshifts (\citealt{Fragos_2013_reionization}) -- and for understanding potential supermassive black hole (SMBH) growth rates in the early universe.
A possible means for measuring the current mass loss rate is via the absorption lines, which are indicators of quasi-relativistic winds (see \citealt{pintoResolvedAtomicLines2016}), requiring observations at $<2$ keV.
Many ULXs are associated with optical or radio bubbles \citep{Pakull2002, soriaBubble2021}, which are sometimes as large as 400\,pc in diameter and mechanically inflated \citep{Gurpide2022}, allowing an estimate for the integrated kinetic energy released by the winds.
The mass inflow rate can be crudely estimated from soft X-ray spectra (see \citealt{Poutanen0705}) or, should the mass transfer dominate the angular momentum distribution within the system, from the change in the orbital period (e.g., see \citealt{bachettiOrbitalDecayM822022}, who estimate the mass transfer rate from a decrease in the orbital period of \Mtwo through pulsar timing).

Winds in XRBs are commonly identified via absorption lines in the Fe K band from hot plasmas (typically Fe\,{\small{XXV-XXVI}}), and in the soft band from a forest of H- and He-like ions of several elements from C to Fe (for a review see \citealt{NeilsenDegenaar2023XRBhighres}).
Initial searches for Fe\,{\small{XXV-XXVI}} in ULXs did not result in detections (see, e.g., \citealt{waltonExtremelyLuminousVariable2013}), most likely due to their soft spectra or high ionisation state of the gas
(for a review on ULX winds see \citealt{PintoWalton2023ULXwindsrev}).
Evidence of atomic lines in low-resolution ULX X-ray spectra were found in the form of residuals around 1 keV with respect to the continuum model through high-count-rate \xmm EPIC spectra \citep{Stobbart2006, 2014Middleton, 2015bMiddleton}. Later on, high-spatial-resolution instruments onboard \chandra showed that the residuals were associated with the ULX itself rather than the interstellar medium in the host galaxy \citep{Sutton2015}, which was supported by the anticorrelation between their strength and the ULX X-ray spectral hardness \citep{2015bMiddleton}. The first unambiguous proof of winds was obtained through long observations (300-500 ks) with the high-spectral-resolution reflection grating spectrometer (RGS) onboard \xmm. In particular, \cite{pintoResolvedAtomicLines2016} discovered a forest of emission lines (at rest-frame laboratory wavelengths) and absorption lines (mainly O\,{\small{VII-VIII}} and Ne\,{\small{IX-X}}), blueshifted by $\sim 0.2c$ in the two well-known ULXs NGC~1313 X-1 and NGC~5408 X-1, in excellent agreement with the predictions of winds driven by strong radiation pressure in super-Eddington accretion disks \citep{Ohsuga2011, Jiang2014}. Modelling of these lines permits an estimate of the kinetic power within the wind (affected by the uncertainties in the covering fraction), which matches well with the luminosities required to inflate the supersonically expanding ULX bubble nebulae (detected primarily in the optical and showing broad lines, see e.g. recent work with the MUSE camera on the Very Large Telescope by \cite{Gurpide2022}).

Features indicating a wind have been found in the majority of ULX RGS spectra with $\gtrsim1,000$ net counts \citep{Pinto2017,kosecEvidenceVariableUltrafast2018,Pinto2021}, as well as in the \chandra gratings observation of the transient Galactic super-Eddington NS, Swift J0243.6+6124\citep{vdEijnden2019}. The luminosity of the plasma producing the emission lines varies slightly over time and is very large ($L_{\rm X}\sim10^{38}$ erg\,s$^{-1}$; \citealt{Pinto2020a}), orders of magnitude higher than in Eddington-limited Galactic XRBs \citep{Psaradaki2018}. Absorption lines at 8.6-8.8 keV from the hotter Fe K component of the wind were finally found (albeit at lower significance) in a few ULXs with hard spectra\citep{waltonIronComponentUltrafast2016,brightman56KeVAbsorption2022}. The emission and absorption lines in the soft band (0.3--2 keV) have equivalent widths similar to those of Galactic XRBs ($EW \sim 5-10$\, eV, see \citealt{Kosec2021}). The Fe K absorption features, albeit rarer, appear deeper ($EW \sim 50-200$\, eV) and similar to the most extreme wind phases observed in Galactic XRBs (which have velocities of $\sim 0.05c$; e.g., \citealt{King2012winds}).

The very low detection rate of Fe K lines arises as a combination of low effective area in current instruments, high background, and low ionic fractions. The latter is a consequence of the soft radiation field of ULXs (e.g., \citealt{PintoWalton2023ULXwindsrev}). However, in ULXs likely seen at low inclinations (e.g., the hard ultraluminous, or HUL, systems of \citealt{Sutton2013} with spectral slopes $\Gamma < 2$), the continuum is strong, likely overwhelming the lines in the soft band (e.g. \citealt{2015bMiddleton}). The Fe K features might then be detected, as the irradiating continuum is hard, as long as the gas is not overionised in the wind cone and the instrumental background is sufficiently low at 7-9 keV.

Although absorption lines have proven elusive in HUL sources (such as Holmberg IX X-1 and most pulsating ULXs), should such features be detected, we could start to understand a) the 3D structure of the wind, and b) the actual mass outflow rate (and true kinetic power). The latter has important cosmological consequences for the net growth rate of SMBHs \citep[e.g.][]{volonteriRapidGrowthHighRedshift2005,volonteriCaseSupercriticalAccretion2015} and the overall role which ULXs play in their host galaxies, particularly at the peak of star formation \citep[e.g.][]{prestwichUltraluminousXRaySources2015}. \hexp will dramatically improve our ability to search for the Fe K wind component in ULXs thanks to its broader energy band (needed for modelling the continuum, which is crucial to search for shallow spectral features, see Fig.~\ref{fig:CRSF_lines}), improved sensitivity compared to \xmm and \nustar (for maximizing total source counts), and its narrower PSF (since ULXs tend to reside in crowded regions).

\subsection{What is the content of the ULX population?}
One of the open questions regarding the nature of ULXs is the prevalence of NS accretors and the strength of their magnetic fields.
The detection of fast ($\gtrsim$ 1~Hz) pulsations unequivocally identifies a ULX as a NS, and pulsar timing can be used to indirectly estimate the magnetic field (e.g., \citealt{ghoshAccretionRotatingMagnetic1979, wangLocationInnerRadius1996}).
The detection of cyclotron lines in X-ray spectra allows us to tackle these two problems simultaneously, as it enables us to identify the presence of a NS where pulsations may be undetectable or absent (\citealt{brightmanMagneticFieldStrength2018}), while simultaneously offering an indication of the magnetic field strength (\citealt{Middleton2019M51, waltonPotentialCyclotronResonant2018}). To date, only two potential CRSFs have been reported in extragalactic ULXs (\citealt{brightmanMagneticFieldStrength2018, waltonPotentialCyclotronResonant2018}), and one Galactic super-Eddington source, Swift J0243+6124 (\citealt{Kong2022_CRSF}). The broadband coverage of \hexp -- and its higher sensitivity above 10\,keV (where these lines are typically detected in NS XRBs; \citealt{Staubert2019}) -- will allow us to increase the population of identified NS-ULXs by detecting CRSFs (see Section~\ref{sub:crsf}).

At the moment, due to the limited amount of high quality data, the sample of well-studied ULXs is very small (a few tens, compared to the thousands known; \citealt{waltonMultimissionCatalogueUltraluminous2022,traninStatisticalStudyLarge2023}), restricting our ability to perform deep searches for pulsations and CRSFs, and hence preventing us from assessing the fraction of NSs in the ULX population.
Given the energy dependence of the pulsing component (e.g. \citealt{Brightman2016_pulsedcomponent}) and the potential for the CRSF to be located at a wide range of energies (depending on the magnetic field strength), broad-band observations, especially at high energies, are crucial.

A similar energy dependence also characterizes other forms of variability in addition to pulsations, such as quasi-periodic oscillations (QPOs), which can in principle be used to classify accreting objects \citep{lewinReviewQuasiperiodicOscillations1988,belloniAtlasAperiodicVariability1990,vanderklisQPOPhenomenon2005}.
ULXs are known to show QPOs \citep{atapinUltraluminousXraySources2019}, some of which match the phenomenology observed in other accreting systems, e.g. QPOs that have been detected over a broad-band (like M82 X-1's 50--300 mHz QPO, see e.g. \citealt{2021ElByad}) are more significant at higher energies ($\gtrsim10$\,keV). The broadband spectro-temporal study of a rich sample of ULXs will help single out possible candidate black hole ULXs (e.g. \citealt{wolterXRayLuminosityFunction2018}).
In particular, it will be crucial for studying the brightest ULXs, also referred to as hyperluminous X-ray sources (HLXs;\citealt{farrellIntermediatemassBlackHole2009,mackenzieHyperluminousRaySource2023}, still considered among the best candidates to host intermediate-mass black holes (IMBHs) although are rare and mostly found in very distant galaxies.

\subsection{What drives the changes in ULX brightness/spectral shape?}
ULXs have been observed to change dramatically in brightness on timescales of days, weeks, and months (e.g. \citealt{walton78DayXRay2016, gurpideLongterm2021,gurpidecycle2021}). Possible causes of this variability include
precession of the disk and wind (e.g. \citealt{Pasham_M82_precessing_disc, 2015bMiddleton, Luangtip2016_HoIX,amatoUltraluminousXraySource2023}) driven by a variety of processes (see the discussion in \citealt{Middleton_2018_Lense_Thirring, Middleton_2019_Accretion_plane}), changes in accretion rate (for instance by the launching of a thermal wind at large radius \citealt{Middleton2022_thermalwinds}, or by modulating the mass transfer rate in a radiatively driven stellar wind), or the onset of a propeller state associated with a NS close to spin equilibrium. The latter has been extensively searched for (\citealt{2018_Earnshaw_propeller_MNRAS.476.4272E, Song2020_propellor}), with the strongest evidence seen in the changing spin period of NGC~5907 ULX-1 (\citealt{Fuerst2023}). With a well considered monitoring strategy, it should also be possible to identify propeller states accompanying smaller changes in luminosity by isolating changes in the spectrum at high energies (see \citealt{middletonPropellerStatesLocally2023}), provided that a broad energy bandpass is available. Determining the mechanism behind the long timescale changes in ULXs then requires several key ingredients: a) high energy coverage and sensitivity to locate and track pulsations, b) low energy coverage and sensitivity to explore how the wind responds to changes in the broad band spectrum (e.g. \citealt{2015bMiddleton, Pinto2020b}), and c) broad simultaneous coverage at high and low energies in order to locate the hard component switching off (whilst the soft emission remains stable), which is an indicator of a likely propeller transition, and more generally to detect spectral transitions \citep{amatoUltraluminousXraySource2023}.

\section{Simulations}

To demonstrate the improved capabilities of \hexp compared to the current primary instruments for ULX studies, \xmm and \nustar, we performed a detailed \hexp simulation of NGC~253, a Galaxy with a large population of bright XRBs and ULXs, using the SImulation of X-ray TElescopes (SIXTE) software package \citep{dauserSIXTEGenericXray2019}, with the procedure described in Section~\ref{sec:sixteproc}.
This simulation provides precise estimates of the source and sky count rates, and the instrumental background, to set up reliable estimates for the detection and temporal and spectral characterization of sources in crowded fields.
These estimates were refined through follow-up simulations, using the count rates and background from the SIXTE simulations as inputs.

All the simulations presented in this paper were produced with a set of response files that represent the observatory performance based on current best estimates as of Spring 2023 (see Madsen et al. in prep.\footnote{Official response files will be released after the submission of the proposal}). The effective area is derived from a ray-trace of the mirror design including obscuration by all known structures. The detector responses are based on simulations performed by the respective hardware groups, with an optical blocking filter for the LET and a Be window and thermal insulation for the HET. The LET background was derived from a GEANT4 simulation \citep{Eraerds2021} of the WFI instrument, and the HET background was derived from a GEANT4 simulation of the \nustar instrument; both simulations assume an L1 orbit for \hexp.
\xmm and \nustar simulations were also performed for comparison, using the official set of response files distributed with SIXTE\footnote{They can be downloaded at \url{https://www.sternwarte.uni-erlangen.de/sixte/instruments/}}.

\subsection{End-to-end simulations of an NGC~253-like Galaxy}\label{sec:sixte}

The \chandra Source Catalogue v.2.0 \citep{evansCHANDRASOURCECATALOG2010,evansChandraSourceCatalog2020} contains 1583 ULXs \citep[following the criteria from][]{waltonMultimissionCatalogueUltraluminous2022}.
Among these ULXs, 14\% have other X-ray sources within 5\arcsec, 45\% within 20\arcsec, and 64\% within 40\arcsec. The vast majority of the ULXs appear close to the nuclear region of their host galaxy, particularly for the ULXs residing in more distant galaxies.
In these conditions, improvements in the angular resolution, whilst not degrading the effective area, timing capabilities, and hard X-ray response, are key for studying ULXs.

ULXs are mostly extragalactic objects; some galaxies, such as M51 (7.2\,Mpc; \citealt{terashimaLuminousXRaySource2004}), NGC~253 (3.5\,Mpc; \citealt{wikSpatiallyResolvingStarburst2014a}), and the Cartwheel Galaxy (150\,Mpc; \citealt{wolterXRayLuminosityFunction2018}), are known to contain a relatively large number of bright accreting sources and ULXs, and so far, only \chandra is able to fully resolve their positions. It is possible that a population of elusive, highly absorbed ULXs \citep[e.g.][]{luangtipDeficitUltraluminousXray2015,westLargeDeficitHMXB2023}, has still escaped detection (or identification as ULXs), awaiting high-energy response (and an ability to spatially separate them).

With its high sensitivity and angular resolution, \hexp will be able to detect and characterize the spectra of many new ULXs. Simulations indicate that the effective area and background of \hexp will allow us to not only identify ULXs from their characteristic spectra, but also detect important spectral features such as emission and absorption lines associated with strong winds, or cyclotron resonance features that allow for an estimate of NS magnetic fields (see Figure \ref{fig:CRSF_lines}).
Moreover, these capabilities are key for detecting pulsations from pulsars in crowded environments.
We demonstrate this in the Sections below.

\subsubsection{SIXTE setup}\label{sec:sixteproc}
For complete, end-to-end simulations, we used the SIXTE software package \citep{dauserSIXTEGenericXray2019}, a Monte Carlo simulation toolkit for X-ray astronomical instrumentation.
This software is able to take into account very accurately the source properties, including simulating complicated fields-of-view with diffuse emission, and characteristics of the optics and detection system of the mission.
The SIXTE team worked with the leads of the \hexp mission to provide up-to-date instrument specification files for \hexp.
The SIXTE website also contains hardware specifications for a number of existing and future missions that allow an easy comparison between them.

To run a simulation with SIXTE, we first generate a source input (SIMPUT-format) file containing: 1) an image of the diffuse emission; 2) its spectral shape defined as a \texttt{Xspec}-format xcm file; and 3) a list of point sources, having different spectral shapes and timing properties, including pulsations and/or aperiodic variability.
Through the script \texttt{run\_sixt}, we then generate realistic simulations for \hexp, \nustar, and \xmm.

\subsubsection{Galaxy and source parameters}
We set up a SIMPUT file using a map of the diffuse emission of NGC~253, and 70 point sources\footnote{The galaxy does not have a bright AGN in the center.
Such a source could, in some cases, dominate the emission close to the nucleus, and would be a source of confusion for the detection of ULXs for all instruments, plausibly more so for \xmm and \nustar, that have a broader PSF.}.

The map of diffuse emission was obtained by removing all point sources from a deep \chandra map of the field. We assigned to it a thermal plasma spectrum (based on \citealt{wikSpatiallyResolvingStarburst2014a}; see the details in \citealt{lehmerHighEnergyXray2023}).
The total 0.5--7\,keV flux of the diffuse emission was $9.6\cdot 10^{-12}$\,\fluxcgs.
The 70 point sources selected were the 70 brightest sources from the \chandra catalogue.
We only conserved the position and flux between 0.5 and 8 keV, but otherwise we changed the spectral and temporal properties of each source, as follows:
\begin{itemize}
\item each source had a ULX-like cutoff-powerlaw spectrum, with random parameters uniformly distributed in intervals deemed plausible by comparison with the spectral characteristics of known ULXs (e.g. \citealt{pintorePulsatorlikeSpectraUltraluminous2017}): cutoff $E_{\rm cut}$ between 1 and 10 keV, and $\Gamma$ between -2 and 0;
\item pulsations with a pulsed fraction between 20\% and 100\% (NGC~300 ULX has $\sim$80\% pulsed fraction over the full band, see e.g. \citealt{carpanoDiscoveryPulsationsNGC2018,vasilopoulosNGC300ULX12019}), a pulse period between 1\,ms and 100\,s, and a pulse profile described by a Von Mises distribution with $\kappa$ between 1 and 5.
\end{itemize}

\begin{figure*}[htb]
    \centering
    \includegraphics[width=\linewidth]{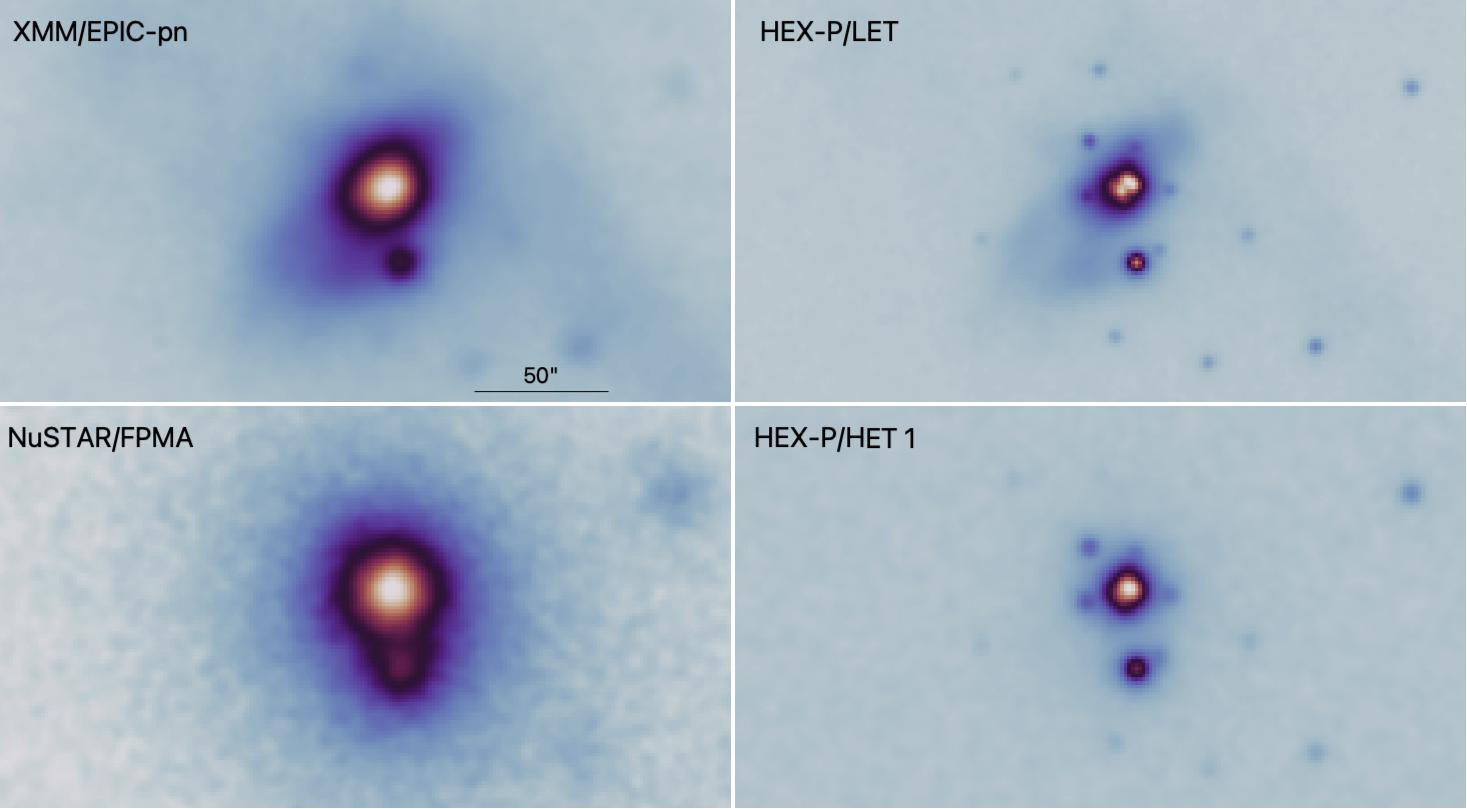}
    \caption{1-Ms Simulation of a NGC~253-like starburst galaxy showing the various components including diffuse hot gas and XRBs. Here we compare the performance of \xmm and \nustar (compare to \citealt{pietschXMMNewtonObservationsNGC2001,wikSpatiallyResolvingStarburst2014a}) with the instruments onboard \hexp, using the procedure described in Section~\ref{sec:population}.
    \hexp is able to resolve most of the bright X-ray sources in nearby galaxies, enabling a detailed study of both the source and diffuse emission.
    In particular, \hexp is able to resolve and study in detail all ULXs, including the ones with known notable features (pulsations, lines, eclipses; see text), with better sensitivity and equivalent-or-better energy and timing resolution compared to \xmm and \nustar}
    \label{fig:population}
\end{figure*}

We then ran full SIXTE simulations for \hexp, \nustar, and \xmm to produce images such as those shown in Fig.~\ref{fig:population}.
To analyze the simulations and estimate count rates from the sources and various components of the background, we used circular source regions centered at the source position with a radius corresponding to half of the half-power diameter of the point spread function used by SIXTE for the simulation.

For each source region, we used the internal description of data in SIXTE to estimate how many photons came from the source $n$ (\texttt{SRC\_ID == n}), how many from the instrumental background (\texttt{SRC\_ID == -1}) and how many from the sky background (\texttt{(SRC\_ID != -1)\&(SRC\_ID != n)}).

\label{sec:population}

\subsection{Pulsar simulations}\label{sec:pulsar}
\begin{figure}
    \centering
    \includegraphics[width=\linewidth]{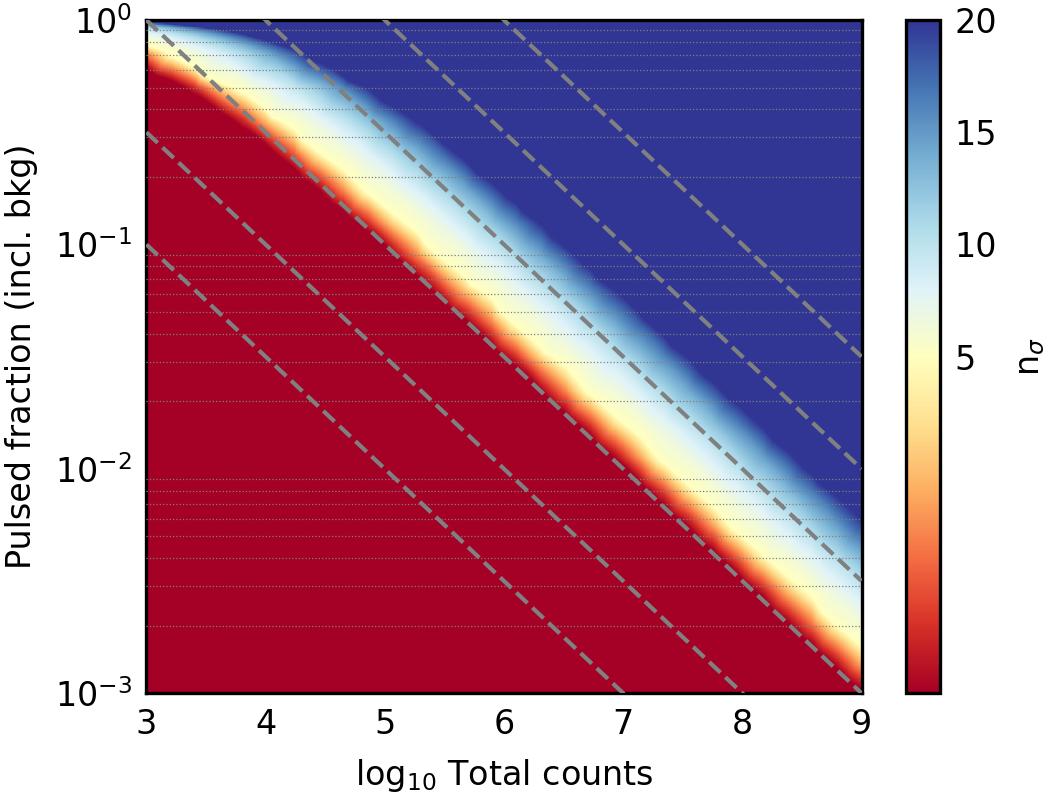}
    \caption{Comparison of the single-trial detection significance (in number of sigma) of a quasi-sinusoidal pulsed profile with a $Z^2_n$ search, given a total number of counts and the pulsed flux fraction (expressed as the counts in the pulse profile divided by the total counts).
    The sensitivity degrades rapidly when the pulsar is not dominating the emission, and a factor of $\sim 5$ lower in pulsed fraction can make a pulsar undetectable. Dashed lines show the expected dependence from a law that scales like Eq.~\ref{eq:rms}.}
    \label{fig:detection}
\end{figure}

Assuming that one has averaged enough periodograms to yield normally-distributed powers, \citet{lewinReviewQuasiperiodicOscillations1988} estimates the significance of the detection of a broad feature against a Poisson background (e.g. a high frequency QPO) in a periodogram as
\begin{equation}\label{eq:rms}
n_\sigma = \frac{1}{2} I r^2 \sqrt{\frac{T_{\rm obs}}{\Delta \nu}}
\end{equation}
where $I$ is the total count rate (sky, instrument, source), $r$ the rms fractional variability of the QPO over the total flux, $T_{\rm obs}$ the observing time, and $\Delta\nu$ the width of the feature.
This means that the observing time necessary to detect a QPO scales with the inverse \textit{fourth power} of the rms.

Pulsar searches are usually performed with single periodograms, or proxies of them such as the $Z^2_n$ statistic \citep{buccheriSearchPulsedGammaray1983} or the Epoch Folding search \citep{leahySearchesPeriodicPulsed1983a}, for which one cannot assume normality of powers. However, it can be shown that a similar dependence on the rms is also valid in this case (\fref{fig:detection}).
Therefore, in order to perform a sensitive search for pulsations, the key ingredient is reducing all components of the background so that the rms amplitude is reduced to the intrinsic source rms.
When pulsars are not dominating their fields-of-view, raw sensitivity of an instrument is less important than its ability to separate the target from the other sources.
ULXs are borderline in this respect, as they are often relatively bright sources in their host galaxies, but being extragalactic, it is common to observe other ULXs or an active galactic nucleus (AGN) separated by less than a few arcminutes.
The vast majority of known ULXs in nearby galaxies can be confidently separated by \hexp (see Sec.~\ref{sec:sixte}).
In addition, their hard X-ray tail, which is where the pulsed emission is strongest, is covered by the energy range where \hexp is most sensitive.
With the typical parameters of ULXs, \hexp can detect pulsations in a fraction of the time it takes \nustar or \xmm, and \hexp can even outperform timing-dedicated, non-focusing instruments, which may have much larger effective areas but far poorer angular resolution.

\subsubsection{Simulation setup}
To simulate time-variable fluxes, we use inverse transform sampling.
By appropriate normalization, a (positive definite) time series can be considered a probability density function.
We first calculate its cumulative distribution function (CDF) by simple integration.
Then, we generate random values of the CDF by extracting a series of values uniformly distributed between 0 and 1.
We then invert this CDF by calculating (through any interpolation algorithm) the time values corresponding to that value of the CDF.
This can be applied to a variable aperiodic light curve, for example generated through the \citet{timmerGeneratingPowerLaw1995b} method, or to the phases of a pulsar given its pulse profile.
If we start from pulsar phases, further corrections to these phases can be made to accommodate spin derivatives, with the usual Taylor series expansion $\Delta\phi=\dot{\nu}t + 1/2\ddot{\nu}t^2 +\dots$.
We can further generate integer pulse \textit{number} values, sum them to the pulse phases, and transform them into times by multiplying by the pulse period. These methods are already implemented in the \texttt{stingray} \citep{huppenkothenStingrayModernPython2019} and \texttt{HENDRICS} \citep{bachettiHENDRICSHighENergy2018} packages that we use for standalone timing simulations, while the \texttt{SIXTE} package takes care of end-to-end simulations \citep[see below]{dauserSIXTEGenericXray2019}.

\subsubsection{A note on frame time}
\begin{figure}
    \centering
    \includegraphics{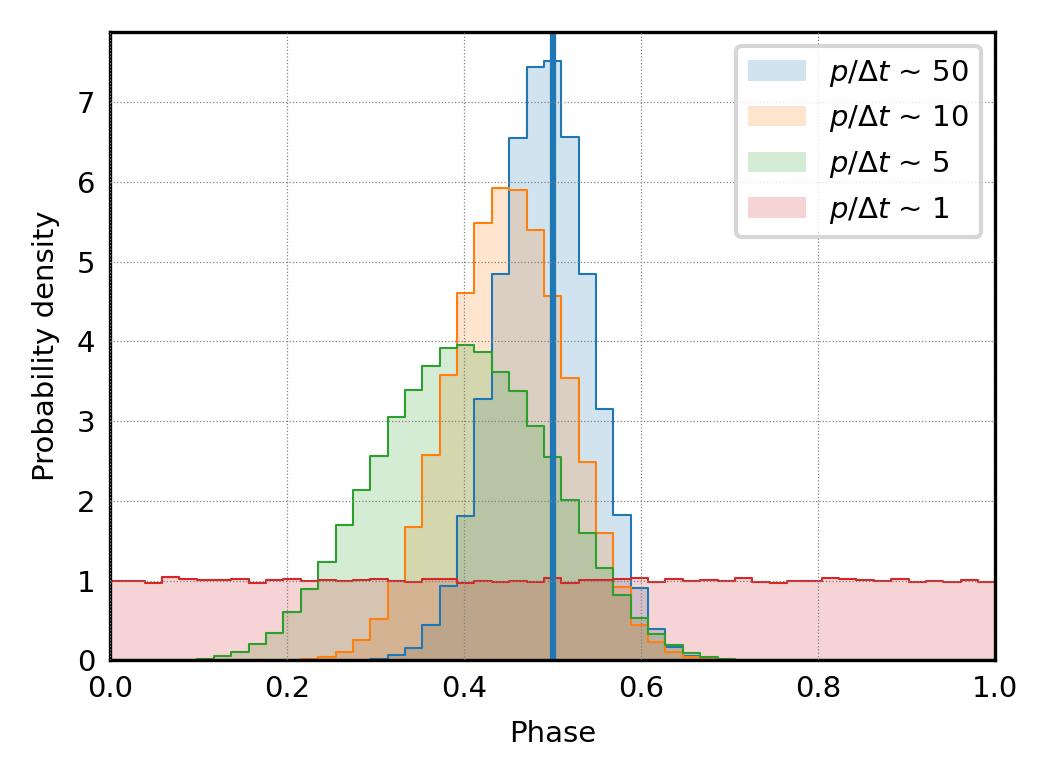}
    \caption{Distortion of a pulse profile when the pulse period $p$ is comparable to the frame time $\Delta t$.
    This is relevant for CCD-based detectors like \xmm's cameras (5.7~ms frame time in small window mode) or \hexp/LET (2~ms frame time).
    }
    \label{fig:frmtime}
\end{figure}

The frame time of CCD-like detectors reduces the sensitivity to pulsations with periods below 10--50 frame times depending on the profile shape (because of the spread of each frame over multiple phases of the pulsations).
Hence, we can expect that the 73\,ms frame time in full window of the EPIC-pn camera onboard \xmm  will greatly reduce sensitivity to sinusoidal pulsations with periods $\lesssim500$\,ms, and sharp pulsations (e.g. some rotation-powered pulsars) with even much longer periods. In Fig. \ref{fig:frmtime}, we generated $\sim$100,000 photons with phases following a Von Mises distribution of concentration ($\kappa=10$), transformed the phases into arrival times by adding an integer pulse number and multiplying by the period, and applied a fixed frame time to the event times.
Then, we folded the profiles using the standard procedure of adding a random number uniformly distributed around $\pm\Delta t/2$ to avoid artifacts\footnote{see, e.g., the \xmm science analysis software documentation, \url{https://xmm-tools.cosmos.esa.int/external/xmm\_user\_support/documentation/sas\_usg/USG/}{https://xmm-tools.cosmos.esa.int/external/xmm\_user\_support/documentation/sas\_usg/USG/}}.
The profiles are clearly distorted, losing sharpness and height as the pulse period approaches the sampling period, until the pulsation is not detectable anymore.

\xmm has special data acquisition modes that reduce the frame time while sacrificing field-of-view.
For example, the ``small window'' mode reaches a $\sim5.7$\,ms frame time by using only $\sim1/40$th of the pixels.
Moreover, the ``timing'' (0.03\,ms) and ``burst'' (7\,$\mu$s) modes also sacrifice angular resolution in order to further increase timing precision \citep{struderEuropeanPhotonImaging2001}.
In contrast, the LET onboard \hexp has a 2\,ms frame time in its standard full frame operation mode, which allows the detection of significantly shorter pulse periods than \xmm, even when conducting surveys or in crowded fields-of-view, making \hexp a game-changer for extragalactic pulsar searches.
The CdTe/CZT-based high-energy instruments such as \nustar and the HET onboard \hexp have a $\sim10\,\mu$s timing resolution, which is more than enough to detect all known periodic and aperiodic phenomena expected in compact objects ($\lesssim2000$\,Hz).
In the following, we assume that all pulsars have periods detectable by all the instruments that are being compared ($\gtrsim1$\,s).

\subsubsection{Pulsar searches}\label{sec:psrsearch}
We carried out two different procedures to search for (and characterize) pulsations in the simulated data described in Sect. \ref{sec:sixte}, one blind search (to evaluate possible confusion of different unresolved candidates) and one directed at known pulsations (using the input periods, and evaluating the increase of detection significance with observing time).

\begin{figure}
    \centering
    \includegraphics[width=\linewidth]{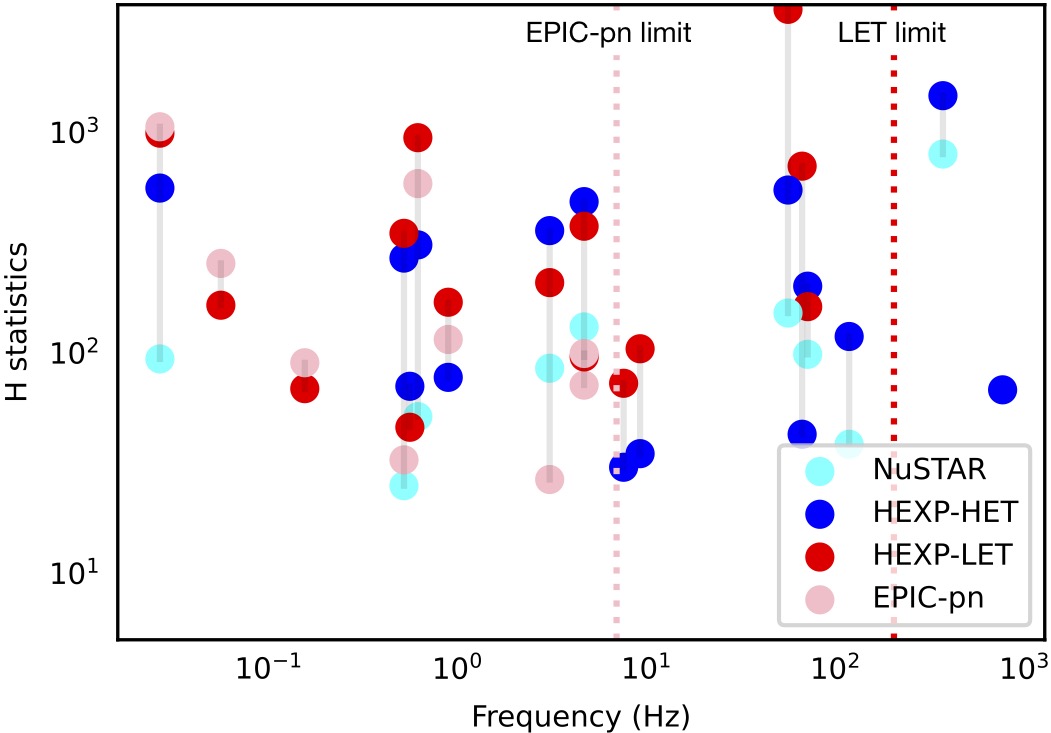}
    \caption{Results of a blind search of pulsations described in Sect.~\ref{sec:psrsearch}.
    Lines connect candidates for the same pulsar in different instruments.
    Some patterns appear: 1) \xmm/EPIC-pn and \hexp/LET have similar performance at low frequencies (the balance depends on pn's slightly larger effective area and LET's lower contamination), while HET is always better than \nustar;
    2) EPIC-pn loses sensitivity above $\sim7$\,Hz due to its frame time, and LET above $\sim$200 Hz.}
    \label{fig:blind}
\end{figure}

For blind pulsation searches, we selected events from all the known source positions, using extraction regions equal to the HPD of each instrument.
We then ran a standard Fourier search with a periodogram, to obtain a list of candidate frequencies \citep{lovelaceDigitalSearchMethods1969}.
Depending on the source position and the instrument used, there was significant contamination between the sources; this made frequency candidates from nearby sources appear in the search of others.
This of course does not represent a problem.
Around each pulsar candidate, we ran zoomed searches using the quasi-fast folding algorithm \citep{bachettiAllOnceTransient2020} as implemented in HENDRICS \citep{bachettiHENDRICSHighENergy2018}. This method allows a fast exploration of the frequency-frequency derivative plane around pulse candidates.
As a metric for pulse detection we used the $H$ test \citep{dejagerPowefulTestWeak1989} modified for binned profiles \citep{bachettiExtendingStatisticsGeneric2021}.
We are mostly interested in the comparison between the instruments, and all parameters affecting the statistics are assumed the same for each\footnote{Here, we are neglecting data loss due to occultation or flaring during the orbit. Taking them into account would advantage \hexp further, because the same exposure would be obtained in shorter observations} (principally the observing time, which increases the frequency resolution of the FFT and hence the number of trials).
We selected a $H$ test statistic of 50 in any instrument as criterion to declare a rough detection.
This resulted in $\sim16$ out of 70 sources being detected by one instrument or the other, and in particular some of the ones in the very center of the galaxy.
For all these detections, we plot the value of the $H$ test calculated from all instruments at the best frequency from the search in Fig.~\ref{fig:blind}.
The results show that \hexp's instruments provide a comparable or higher sensitivity to pulsations compared to \xmm and \nustar.
Moreover, the two combined \hexp instruments yield high sensitivity to pulsations in both soft and hard sources, something difficult to achieve with current instruments without complicated coordination.
Finally, the LET's short frame time allows for sensitive searches of pulsations at frequencies well above \xmm/EPIC-pn's limit.

\begin{figure*}
\centering
    \includegraphics[width=0.9\linewidth]{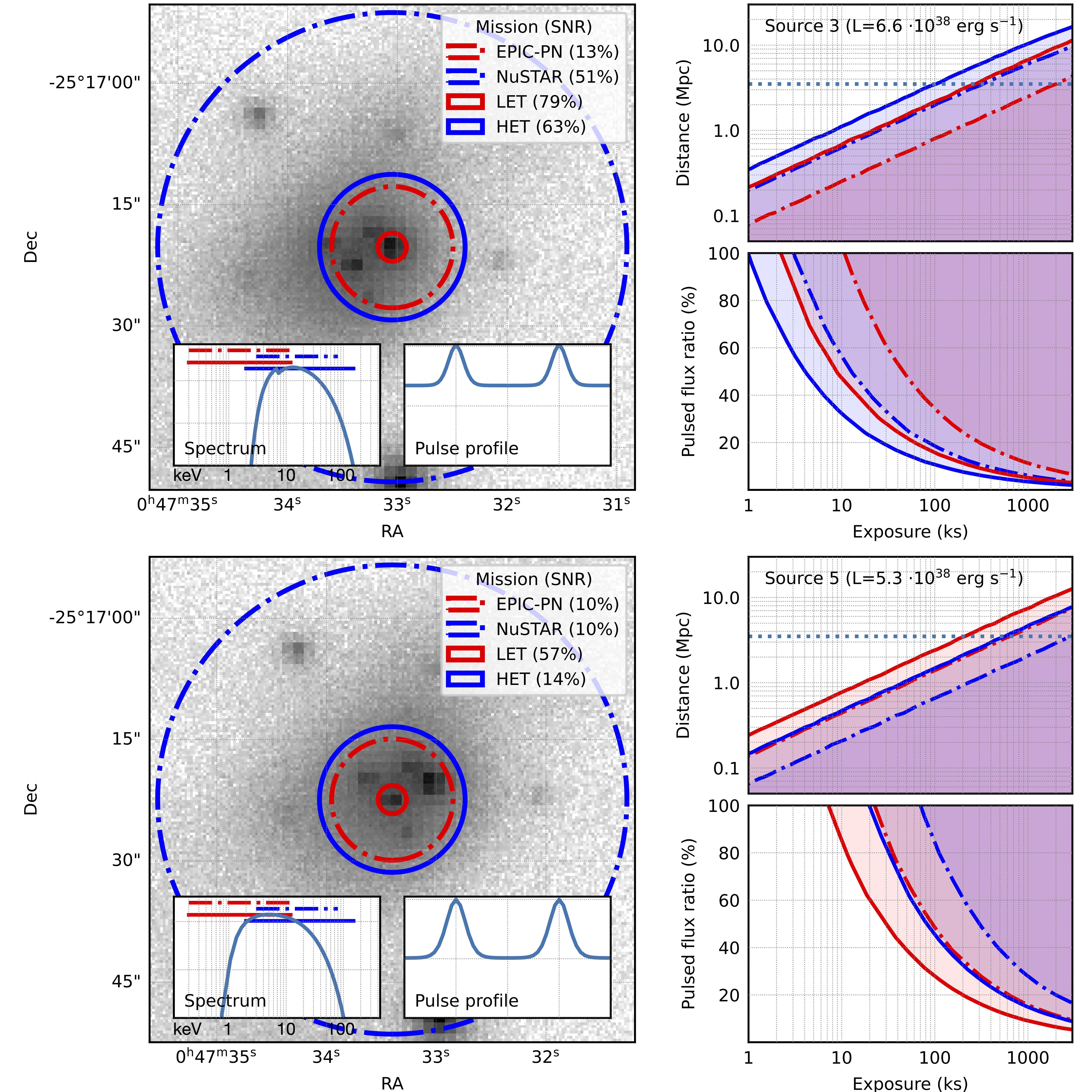}
    \caption{The maps on the left show a 1-Msec LET simulation of the central region of NGC~253 using the same setup as Fig.~
    \ref{fig:population}.
    Each map is centered on one example source in this region, and the two insets show the spectrum and the pulse profile of the source.
    Circles indicate the HPD of each instrument. The two sources are only resolved by the LET.
    The shaded regions on the right of each map show (top) the limiting distance at which we can detect the pulsation (at 5~$\sigma$) at the given PFR with different exposures, and (bottom) the limiting intrinsic PFR detectable at the distance of NGC~253 (3.5\,Mpc, marked in the top plot as a blue dashed line).
    Color coding for all line plots indicates different instruments (the band pass in the first inset).
    \nustar and \hexp/HET include the flux from two modules.
    Percentages in the legend indicate the source to background (all components) flux ratio in the region.
    }\label{fig:confused}
\end{figure*}

To measure the expected detection significance of pulsars injected in the SIXTE simulation, we first evaluated the $H$-test at the injected frequency and the integrated pulse profile over the full simulation, and found the correct number of harmonics $M$ to describe the pulsed profile.
We then estimated the variation in the $Z^2_M$ statistic for different exposures and source distances in the following way:
\begin{enumerate}
    \item We created a grid of exposure times ($E$) and distances ($D$) for the source.
    For each value of $E$ and $D$, we rescaled all sky counts (background and source) with a $E/D^2$ law, while the instrumental background was rescaled by exposure alone.
    \item For each of the rescaled count estimates, we constructed a model pulse profile, as follows: 1) the theoretical pulse profile, including its intrinsic pulsed fraction, was normalized to an integrated value of 1 and multiplied by the number of source counts; 2) we added a background with the expected counts per bin from the sky and the instrument.
    \item Finally, we created 100 realizations of the pulse profile by simulating Poisson-distributed random numbers for each bin of the pulse profile, and we calculated the $Z^2_M$ statistic for each realization.
    We then calculated the false-alarm probability for the average value of the $Z^2_M$ statistic (using \citealt{buccheriSearchPulsedGammaray1983}, as adapted to folded pulsed profiles by \citealt{bachettiExtendingStatisticsGeneric2021}) over all realizations, and expressed it in terms of Gaussian sigma values for ease of interpretation.
    This was then used to form the plots on the right hand side in Fig.~\ref{fig:confused}.
\end{enumerate}

Fig.~\ref{fig:confused} shows the detectability of two simulated bright pulsars (not even at a ULX level of brightness), near the nucleus of a galaxy modeled on NGC~253, with different instruments; these provide good examples of what happens when one looks for pulsations from sources in a crowded field and were selected to have a softer and  harder spectrum respectively.
The detectability of specific pulsars depends on the spectral shape, the fraction of pulsed flux, and the sky and instrumental background.
Therefore, the same intrinsic pulsed flux ratio (PFR)\footnote{Note that with PFR we are comparing the pulsed \textit{flux}, the integral of the AC component of the pulse, to the total flux. The more commonly used pulsed fraction (PF), instead, compares the \textit{amplitude} of the pulsation to the mean or the maximum of the pulse. Our definition is more independent of pulse shape.} can result in very different total PFR (and hence, detectability) in different instruments, depending on sensitivity and angular resolution.
In these two cases (and the same consideration is supported by the blind search as well, Fig.~\ref{fig:blind}), one can see why the instruments onboard \hexp perform equivalently or better than current focussing instruments like \xmm and \nustar, especially if the source is located in crowded fields or even in peripheral regions of the host galaxy.
The angular resolution of \hexp allows the emission from a large number of contaminating sources to be filtered out, improving the PFR and the pulsation sensitivity, even where the effective area is slightly lower (e.g. \xmm/EPIC-pn compared to LET).

\subsection{Spectral simulations}

As discussed in Sect. 1, the X-ray spectral continuum of ULXs can be complex, characterized by up to three main components: a low-temperature ($<1$ keV) thermal component, a hotter 1-3 keV component, which appears broader than the cool one, and a hard component above 10 keV (e.g., \citealt{waltonEvidencePulsarlikeEmission2018}). The latter can appear as a hard excess in the \nustar band, and may be related to Compton up-scattering of the disk emission, which is not included in the simple accretion disk models typically used (c.f. post-processing of general relativistic magnetohydrodynamical simulations of super-Eddington discs, \citealt{Mills2023}).
In some cases it is likely associated with the accretion column of a NS. The cool thermal component is often modelled with a simple disk-blackbody component (e.g.,  diskbb in \texttt{Xspec}) approximating the emission from the outer disk and/or wind. The hotter thermal component has a broader spectrum and can be equally described by a slim disk (e.g., diskpbb in \texttt{Xspec}) with a temperature-radius law $T \propto R^{-p}$ (with $p$ taking values 0.5--0.75, e.g. \citealt{waltonUnusualBroadbandXray2020}) or a thermal Comptonisation component (e.g., nthcomp in \texttt{Xspec}, \citealt{2015aMiddleton}). Finally, when identified with a pulsation, the emission from the accretion column can be empirically modelled with a cutoff-powerlaw (\textsc{cutoffpl}) with slope $\Gamma = 0.59$ and high-energy cutoff $E_{\rm{cut}} = 7.9$\,keV, which is based on the average parameters found for known ULX pulsars via phase-resolved spectroscopy (e.g., \citealt{waltonPotentialCyclotronResonant2018}).

The study of the relationship between the luminosities and temperatures of the thermal components measured in different epochs may be useful in order to obtain valuable information on the evolution and structure of the inner accretion disk. The luminosity-temperature ($L-T$) trends often disagree with the theoretical predictions of sub-Eddington thin discs ($L \propto T^4$) and even show negative slopes consistent with the photosphere of a super-Eddington disk/wind (e.g., \citealt{Robba2021}). In this regard, \hexp will provide an important improvement with respect to current missions thanks to its high effective area and broad energy coverage. In Walton et al. (in prep.), for instance, simulations of Holmberg IX X-1 have been performed showcasing how \hexp will enable accurate estimates of the $L-T$ trends. In particular, it will be possible to distinguish the $L-T$ tracks observed at high and low source fluxes (also seen in NGC~1313 X-1 \citep{waltonUnusualBroadbandXray2020}), which can indicate dramatic changes in the inner accretion flow.

Once a satisfactory description of the continuum is achieved, spectral residuals can be located (almost ubiquitously: \citealt{2015bMiddleton}) and, eventually, searches for narrow lines can be performed. This is typically performed through scans that involve a moving Gaussian through a grid of centroid energies (e.g., \citealt{pintoResolvedAtomicLines2016}). Monte Carlo simulations must be used in order to probe the statistical significance of any putative lines (e.g., \citealt{Kosec2021}). More recently, the exploration of a large parameter space has been employed through grids of plasma models, which account for multiple emission or absorption lines simultaneously. This boosted the detectability of weak lines such as those from ultrafast outflows (e.g., \citealt{kosecEvidenceVariableUltrafast2018, Pinto2021}). The emission lines are primarily produced by Ne K and Fe L ions around 1 keV and the absorption lines from blueshifted O VIII and Ne K / Fe L around 0.8 and 1.2 keV, respectively.  Evidence of Fe K around 8--9 keV has also been found, showing the hottest, fastest and most powerful wind component (e.g., \citealt{waltonIronComponentUltrafast2016,brightman56KeVAbsorption2022}). The emission and absorption lines can be well described with models of plasma in photoionization equilibrium, which is expected in the case of winds irradiated by the ULX continuum. Most work in this regard has made use of the \texttt{Spex} code \citep{kaastraSPEXNewCode1996}, in particular through the \texttt{pion} component, although some work also made use of \texttt{Xstar} \citep{kallmanPhotoionizationHighDensityGas2001,mendozaXSTARAtomicDatabase2021} coupled with \texttt{Xspec} \citep{arnaudXSPECFirstTen1996}. In the latter case, photoionization model grids are computed with \texttt{Xstar} and then loaded into \texttt{Xspec}, whilst in the former, \texttt{Spex} instantaneously fits the continuum, calculates the ionization balance and the model, and fits it to both lines and continuum. An example of \hexp's performance with regards to continuum modelling and line detection is shown in Sect. \ref{sec:wind_features}.

\subsubsection{Simulation setup}
A typical workflow for X-ray spectral simulations involves: 1) defining a theoretical model; 2) defining a count rate $\mathcal{C}$ and observation length $\mathcal{L}$; 3) specifying a set of instrument responses (RMF and ARF); 4) specifying an appropriate background file or a model providing an estimate of the instrumental and sky background contaminating the source region; 5) using a tool that generates a list of $\mathcal{CL}$\footnote{Or better, a random number following a Poisson distribution around that value.} random X-ray energies following a statistical distribution obtained by the convolution of the spectral model with the instrumental responses.

In our case, the last step in the above list is executed with \texttt{Xspec}'s  \texttt{fakeit} command, using the most recent response files available for each mission (see Sect. \ref{sec:wind_features} and \ref{sub:crsf}), following the procedure from the \texttt{Xspec} manual\footnote{\url{https://heasarc.gsfc.nasa.gov/xanadu/xspec/manual/node41.html}} or the equivalent procedure for PyXspec\footnote{\url{https://heasarc.gsfc.nasa.gov/xanadu/xspec/python/html/extended.html}}. In order to simulate  emission or absorption lines from plasmas in different types of equilibria (collisional or photoionisation) we also used the \texttt{Spex} code \citep{Kaastra199spex}, which provides up-to-date atomic data, plasma models and a fitting package\footnote{\url{https://spex-xray.github.io/spex-help/index.html}}.

\subsubsection{Wind features}
\label{sec:wind_features}

\begin{figure}
    \centering
    \includegraphics[width=0.975\linewidth]{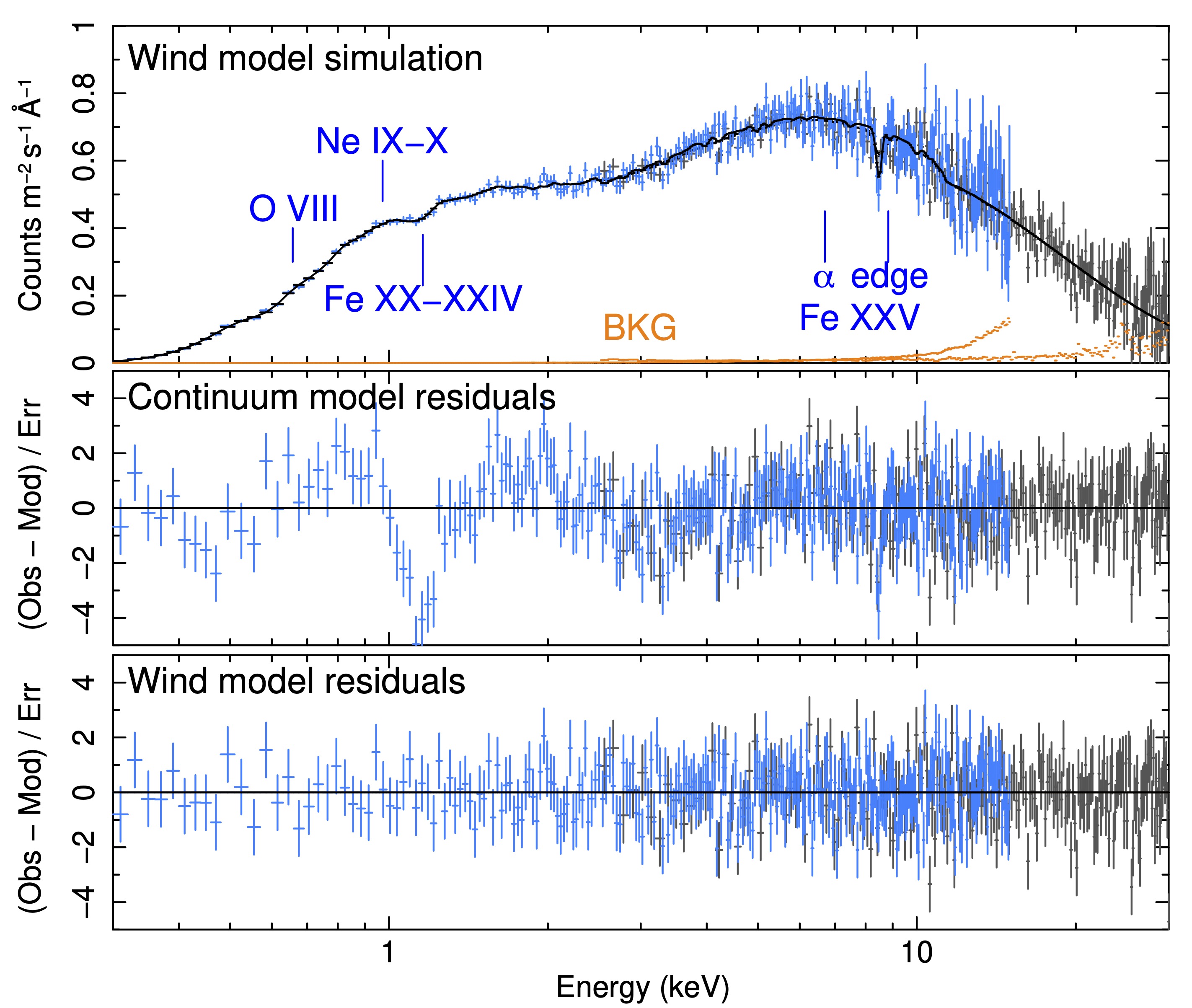}
    \caption{500ks \hexp simulation of ULX NGC~1313 X-1 adopting the best-fit multiphase wind model of photoionised emitting and absorbing plasma (\citealt{Pinto2020b}). The middle panel shows the residuals assuming a continuum-only model (double blackbody plus cutoff powerlaw). The bottom panel shows the residuals to the full wind model. The rest-frame energies of the dominant transitions are labelled. Each component producing emission/absorption lines is detected at $\gtrsim5\,\sigma$.}
\label{fig:wind}
\end{figure}

In order to demonstrate the power of \hexp in studying winds from ULXs by e.g. achieving a highly significant ($\sim5\,\sigma$) Fe K detection, we performed a simulation of the archetypal ULX NGC~1313 X-1, adopting the model described in \cite{Pinto2020b}. The continuum consists of cool (0.2 keV) and hot (2 keV) thermal components and a cut-off powerlaw, the latter dominating above 10 keV with a slope of $\Gamma = 0.59$ and high-energy cutoff $E_{\rm{cut}} = 7.9$\,keV as mentioned above \citep{waltonUnusualBroadbandXray2020}. Starting from previous work on NGC~1313 X-1, we simulated a deep \hexp observation adopting the best-fit multiphase wind model of photoionised emission (at rest) and absorption (blueshifted by 0.2$c$) plasmas (see \citealt{Pinto2020b}). This consists of one {\scriptsize{PION}} emission component and two {\scriptsize{XABS}} absorption components in \texttt{Spex}. The results are shown in Fig. \ref{fig:wind}. The LET and HET simulated spectra in the top panel show a very strong absorption feature in the soft band above 1 keV from the cooler phase of the wind, and a narrow absorption line just above 8 keV from the hot phase of the wind (mainly Fe XXV; c.f., \citealt{waltonIronComponentUltrafast2016}). We also highlight the contribution of the wind by removing the emission and absorption lines from the model and refit the simulated spectra with a continuum-only model (double thermal component plus cutoff powerlaw; see \citealt{waltonUnusualBroadbandXray2020}). Each of the three emission and absorption line components is detected\footnote{As determined through $\Delta\chi^2$, using the affected spectral bins as the number of free trials} at $\gtrsim5\,\sigma$ assuming an exposure of 500\,ks, whilst for the weak Fe K lines confidence levels around $3\,\sigma$ are found with 300\,ks observations, which are significantly shorter than the 700\,ks total \xmm and \nustar observations needed. Moreover, any observation that can detect the Fe K absorption will easily detect the lower energy ($\sim1$ keV) absorption as well since the latter typically requires exposures of a few tens of ks for bright ULXs. There are about 10 ULXs with fluence comparable to NGC~1313 X-1 in the Fe K band and a few dozen within a factor of a few \citep{waltonMultimissionCatalogueUltraluminous2022}. We therefore be confident that \hexp will enable the first meaningful search for Fe K lines in a statistical sample of ULXs, which will place new constraints on the wind geometry, its outflow and kinetic energy, and improve our understanding of the overall accretion mechanism.

\subsubsection{Cyclotron lines} \label{sub:crsf}

Three (potential) CRSF features have been discussed for ULXs/related sources in the literature: an unusually low-energy $\sim$4.5 keV feature in M51 ULX-8 that could be a proton CRSF \cite[seen by \chandra; ][]{brightmanMagneticFieldStrength2018}, an extremely high-energy feature at $\sim$150 keV in the Galactic, Swift J0243+6124 \cite[seen by HXMT;][]{Kong2022_CRSF}, and a potential CRSF at $\sim$13 keV in the ULX pulsar NGC300 ULX1 \cite[seen in a coordinated \xmm + \nustar observation;][]{waltonPotentialCyclotronResonant2018}. However, the detection of the latter is noted to be somewhat tentative, as the data can also be explained by a more complex continuum model with no CRSF, an issue also discussed by \cite{Koliopanos2019} who argued the more complex continuum model may actually be preferred by the data. Part of the uncertainty is the poorly understood continuum and the limited coverage above $\sim$30\,keV offered by \nustar, which makes it an excellent case-study for \hexp.

In order to estimate the exposure time needed to unambiguously detect the putative line in NGC~300~ULX-1, we have simulated spectral models with and without the feature, using the models reported by \citet{Koliopanos2019} (see their Table 1 where they use a combination of two multi-color disk blackbodies and a powerlaw for the high energy emission, referred to as their MCAE model). We focus our simulations on the MCAE model with and without the line, as this gave the highest (Bayesian) evidence among a set of competing models (see their Table 2), but below we also present an alternative case.

We estimated the CRSF detection significance $\alpha$, using the method from \citet{Protassov2002}. To this end, we simulated a single \hexp spectrum (both LET and HET) using the MCAE model with a CRSF line (as reported in \cite{Koliopanos2019}) for exposure times of 50, 100, 150 and 200\,ks. We fitted this spectrum with the null hypothesis model of the MCAE model {\it without} a line and then with the MCAE model {\it with} the CRSF, recording the single value of $\Delta \chi^2_\text{CRSF}$ fit-improvement for each exposure time. We then simulated 1,000 spectra from the best-fit null hypothesis model and fitted these with both models, also recording the fit-improvement in each instance (obtaining the distribution of $\Delta \chi^{2}_\text{null}$). We compared this fit-improvement distribution $\Delta \chi^{2}_\text{null}$ with the $\Delta \chi^2_\text{CRSF}$ value obtained from the \hexp spectrum that included the CRSF, thereby obtaining the probability of observing a $\Delta \chi^2$ fit-improvement as high as that observed in the original \hexp spectrum (i.e. the line detection significance $\alpha$). We repeated the whole process 10 times per exposure to estimate the variations in $\Delta \chi^2_\text{CRSF}$ and $\alpha$ due to Poisson fluctuations only.

\begin{figure*}
    \centering
    \includegraphics[width=0.49\textwidth]{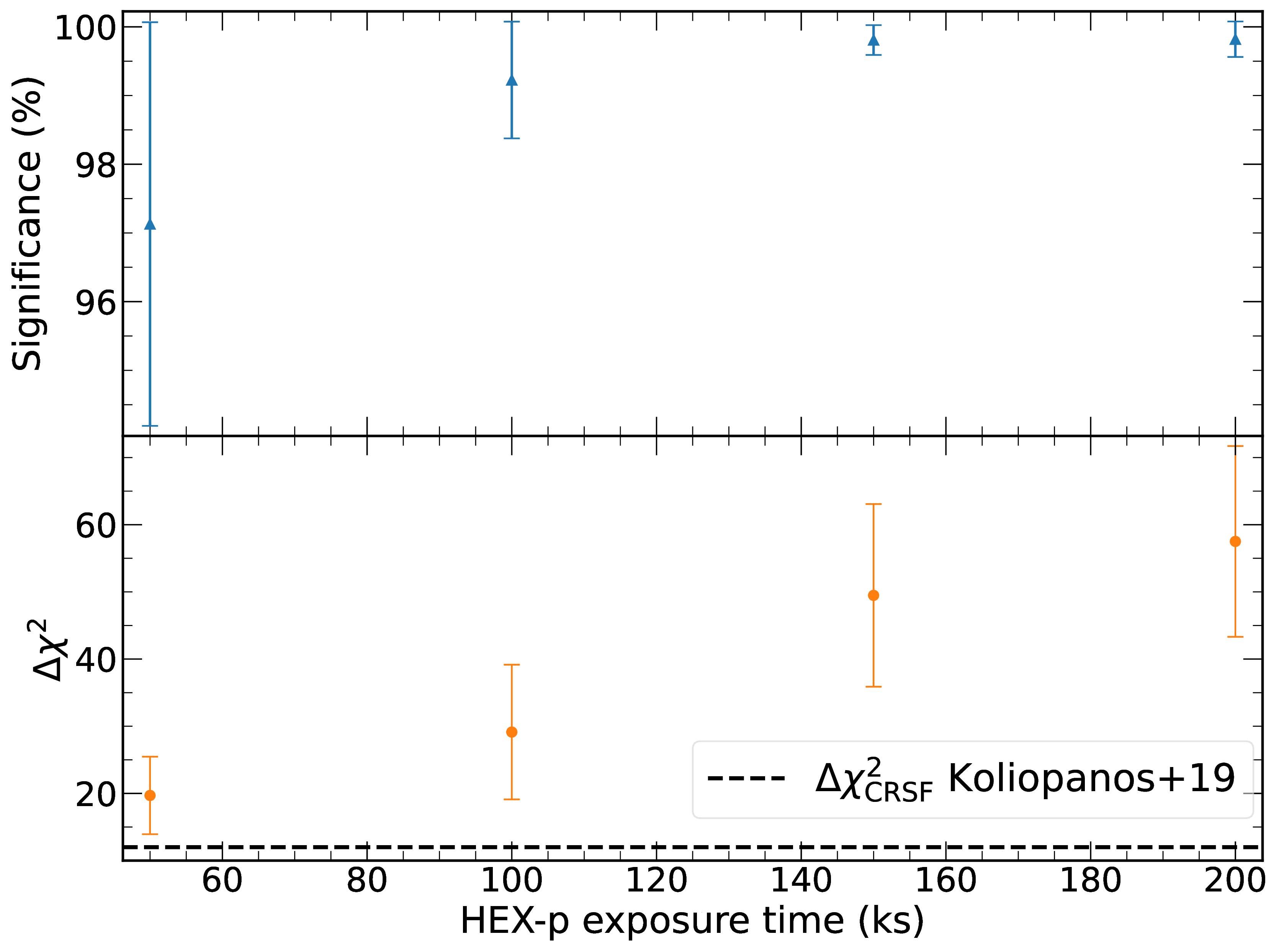}
    \includegraphics[width=0.47\textwidth]{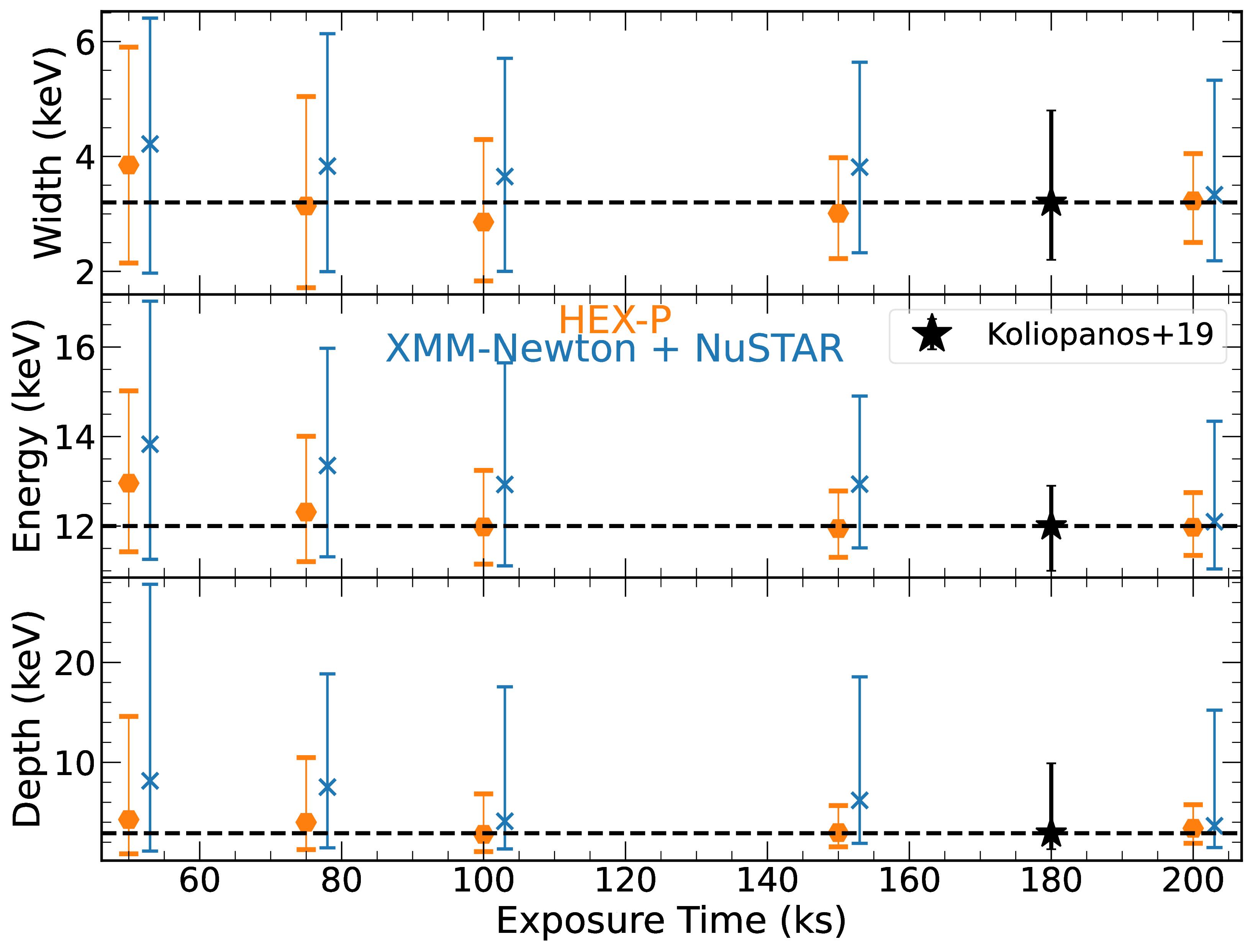}
    \caption{\hexp detection levels and constraints on the putative NGC~300~ULX-1 CRSF. (Left) Line detection significance (top) and $\Delta \chi^2$ fit-improvement (bottom) as a function of exposure time. The error bars represent the spread over 10 realizations (see text for details). The $\Delta \chi^2$ fit-improvement obtained by \citet{Koliopanos2019} using a combined \xmm and ($\sim$140 ks for EPIC-pn) \nustar ($\sim$180\,ks) observation is shown as a dashed line for comparison. (Right) Constraints on the NGC~300~ULX-1 CRSF centroid energy (top) width (middle) and depth (bottom) as a function of exposure time for \hexp (orange) and \xmm + \nustar (blue, values have been shifted slightly horizontally for clarity). The mean best-fit value and mean (90\% level) uncertainty are shown after averaging over 25 spectra per exposure time. The stars indicate the constraints obtained by \citet{Koliopanos2019} for comparison. }
    \label{fig:ngc300_ulx}
\end{figure*}

\begin{figure}
    \centering
    \includegraphics[width=0.49\textwidth]{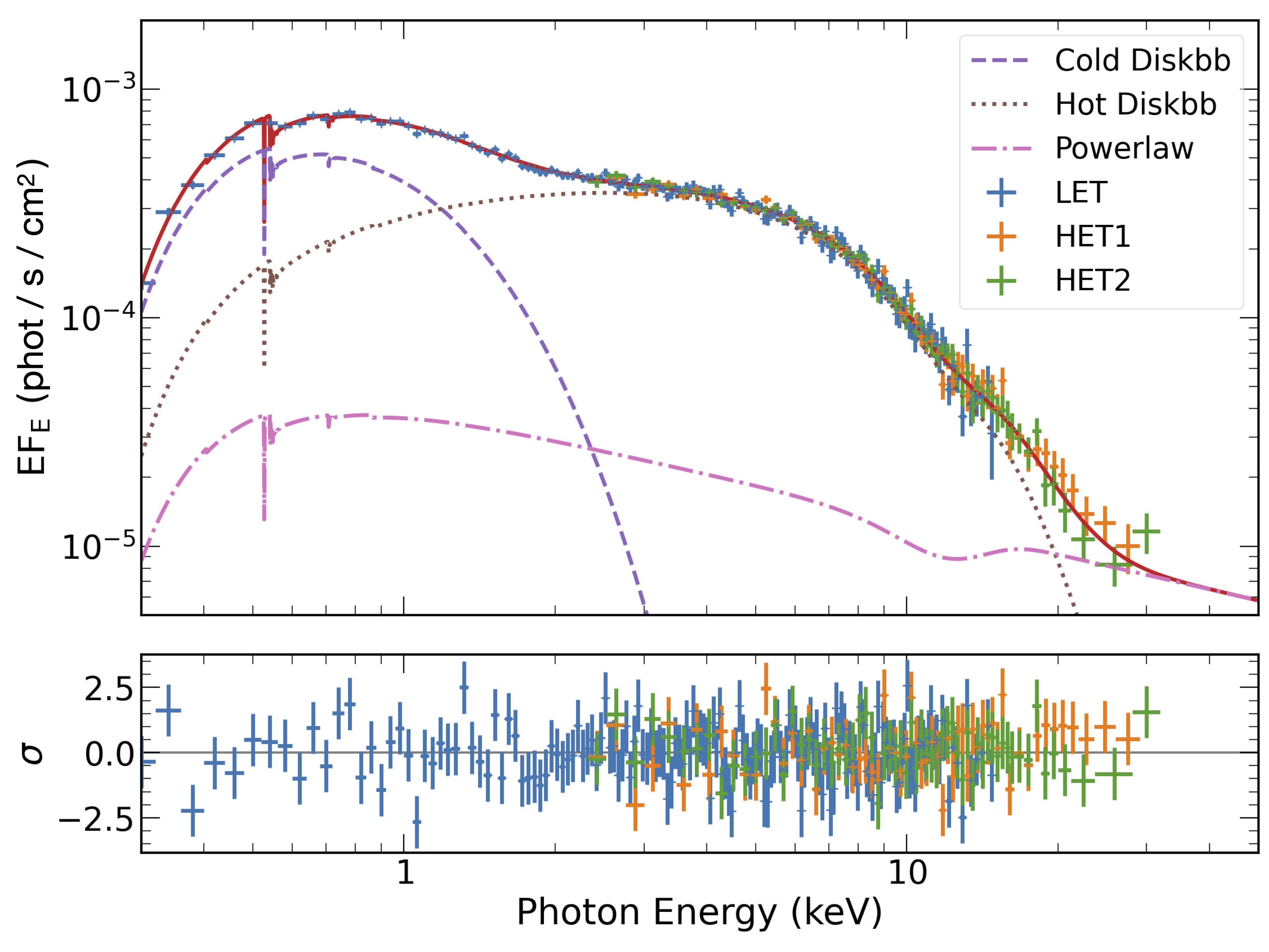}
     \includegraphics[width=0.49\textwidth]{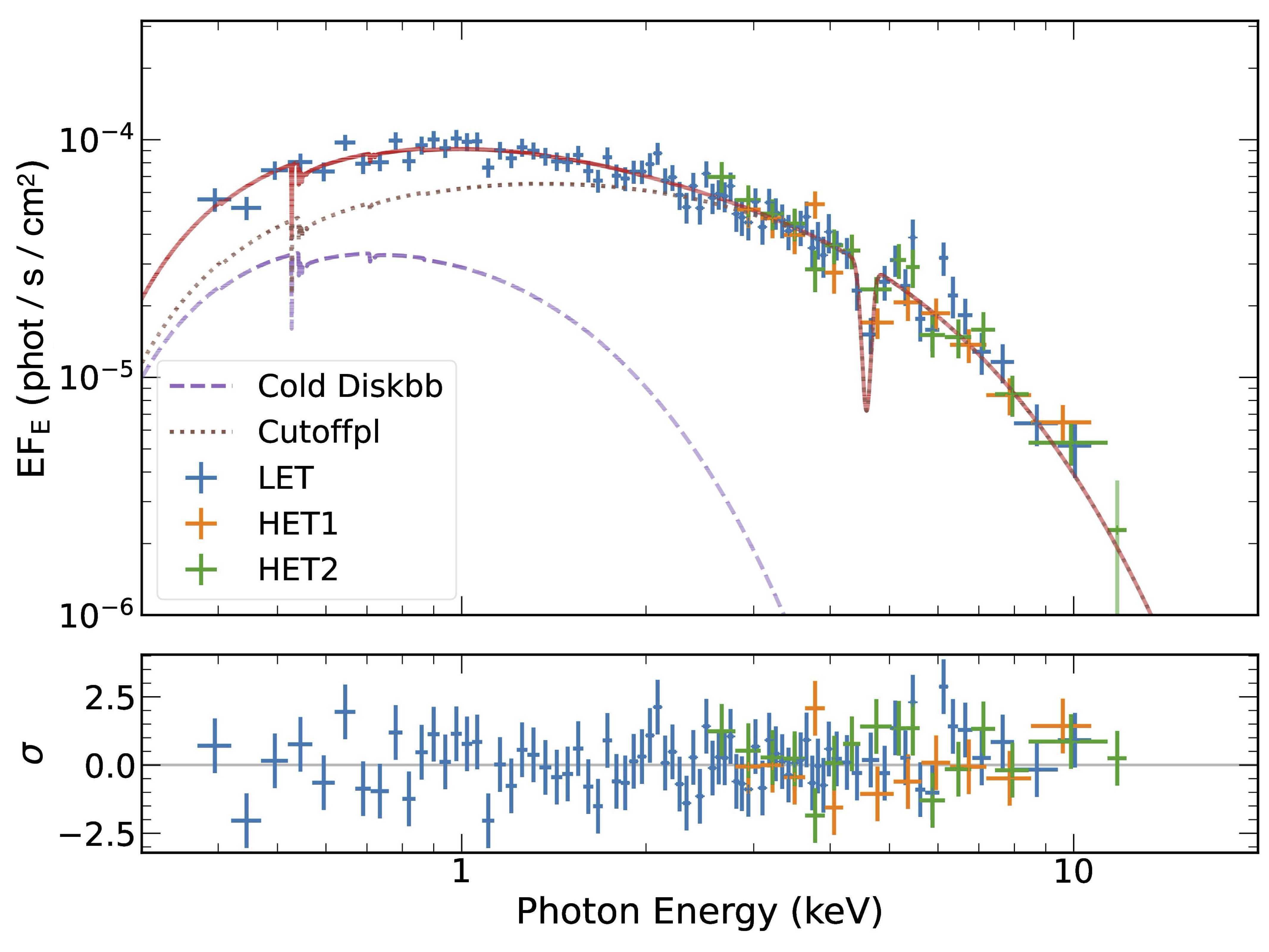}
    \caption{(Top) CRSF line in NGC~300~ULX-1 as observed by \hexp in a 200\,ks observation. (Bottom) CRSF line in M51~ULX-8 as observed by \hexp in a 100\,ks observation.}
    \label{fig:CRSF_lines}
\end{figure}

The results of our simulations are shown in Fig.~\ref{fig:ngc300_ulx}. We see that for a 50\,ks exposure, including the line represents a $\Delta \chi^2_\text{CRSF} \sim 19\pm6$ the 1$\sigma$ uncertainty comes from the spread of the 10 simulations improvement over the null continuum, which translates to a $\sim$97\% (2.2$\sigma$) detection significance, although the high-energy power-law will be marginally detected. Such a $\Delta \chi^2_\text{CRSF}$ is already higher than, or comparable to, the value obtained by \citet{Koliopanos2019}, who reported a $\Delta \chi^2_\text{CRSF} = 12$ improvement based on the existing $\sim$140 ks EPIC-pn and 180\,ks \nustar exposure (see Fig.~\ref{fig:ngc300_ulx} for details). For a 100\,ks exposure time with \hexp, the line is detected at the $\sim 99\%$ ($\sim$2.6$\sigma$) confidence level. Here the line provides a $\Delta \chi^2_\text{CRSF}$ = 29$\pm$10 fit-improvement.
Finally for exposure times greater than 150\,ks, we found the line will be detected comfortably above the 99.73\% (3$\sigma$) level with $\Delta \chi^2_\text{CRSF} \sim 50\pm14$ and $58\pm14$ for 150 and 200\,ks respectively, with the high-energy power-law significantly detected. This final spectrum as seen by \hexp is shown in Fig.~\ref{fig:CRSF_lines}.

We also obtain simulated constraints on the line energy, width and depth, and compare these to those reported by \citet{Koliopanos2019} in Fig.~\ref{fig:ngc300_ulx}. We also simulated \xmm-\nustar observations including all EPIC and both FMPA/B cameras, using responses from a similar epoch to the NGC~300~ULX-1 observations. We note that our latter simulations ignore calibration inaccuracies between \xmm and \nustar detectors, which will only degrade the results presented here, particularly as compared to \hexp. We can see from Fig.~\ref{fig:ngc300_ulx} that \hexp will allow us to obtain tighter constraints on the line properties compared to dedicated \xmm + \nustar observations (about a factor two or even more).
As stated above, one of the main uncertainties in the identification of the CRSF in NGC~300 ULX-1 was the nature of the underlying continuum. \citet{waltonPotentialCyclotronResonant2018} proposed to tackle the problem by performing pulse-phase resolved spectroscopy. In particular, \citet{waltonPotentialCyclotronResonant2018} isolated the pulsed spectrum by extracting data from the brightest and faintest quarter-phases of the pulse cycle of NGC~300~ULX-1 (and obtained the difference spectrum i.e. `pulse-on' - `pulse-off'). By doing so, the uncertainty on the continuum is reduced as the `constant' or non-pulsing component is eliminated. We thus performed additional simulations to examine the ability of \hexp to distinguish between the two different continuum models presented by \citet{waltonPotentialCyclotronResonant2018}.

\begin{figure}
    \centering
    \includegraphics[width=0.49\textwidth]{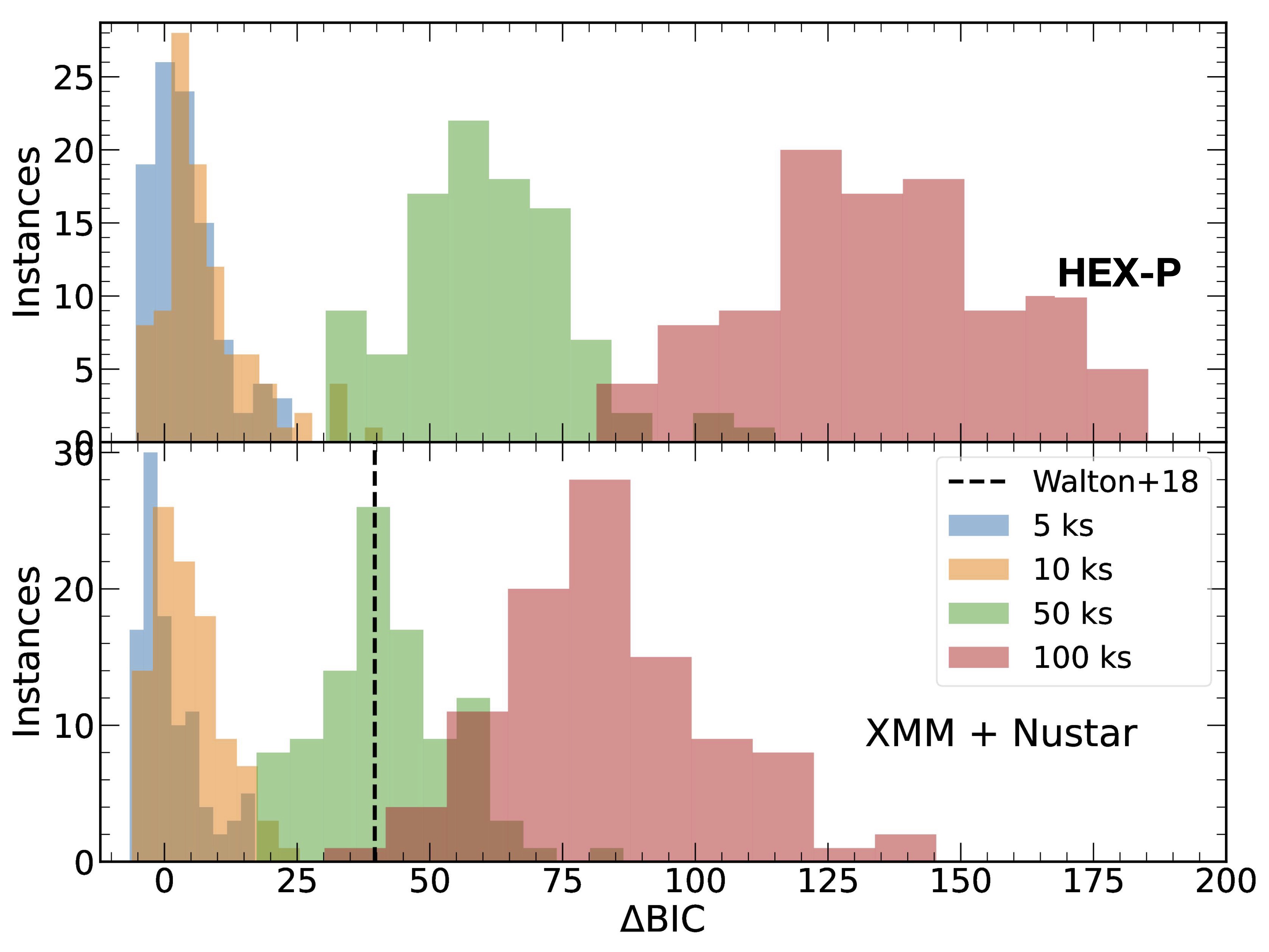}
    \caption{$\Delta$BIC between the two competing models presented in \citet{waltonPotentialCyclotronResonant2018} (a simpler continuum including the CRSF vs a more complex featureless continuum) fitted to the pulsed component of NGC~3000~ULX1 for \hexp (top) and combined \xmm + \nustar observations (bottom) for different exposure times. 100 realizations are shown for each exposure time, as indicated by the colors. For comparison, we also show the value retrieved by \citet{waltonPotentialCyclotronResonant2018} using 140 of EPIC-pn,  190\,ks of combined EPIC-MOS, and 180\,ks of combined \nustar observations (dashed line). We can see that \hexp would be about a factor 1.5 more effective (notably above $\sim$30\,ks) in distinguishing these two different continuum models. Note that the exposure times are given for a quarter of the full pulse cycle and therefore total exposure times would be four times the values shown.}
    \label{fig:ngc300_ulx_BIC}
\end{figure}

We simulated separate (LET and HET) spectra for the high (pulse on) and low (pulse off) phases of the pulse cycle of NGC~300 ULX-1 (each spanning 0.25 in phase) using the models of \citet{waltonPotentialCyclotronResonant2018}, i.e. a combination of two non-pulsing blackbodies and a high-energy {\sc gabs}$\otimes${\sc cutoffpl} ascribed to the accretion column. We subtracted the two and then fitted the pulsed component with the two competing models presented by \citet{waltonPotentialCyclotronResonant2018} (a model containing the CRSF line, {\sc gabs} $\otimes$ {\sc cutoffpl}, and a more complex featureless continuum, {\sc cutoffpl} $\otimes$ {\sc simpl}) and retrieved the difference in Bayesian Information Criterion ($\Delta$BIC; \citealt{schwarz_estimating_1978}) between the two models for 100 simulations per exposure time. From Fig.~\ref{fig:ngc300_ulx_BIC} we can see that \hexp would be about a factor 1.5 more effective (notably above $\sim$50\,ks per phase bin, corresponding to a total 200\,ks) than \xmm \& \nustar in distinguishing different continuum models, therefore reducing uncertainties related to the presence of CRSF lines in the ULX spectra.

The same procedure as above was followed to explore the ability of \hexp to detect the 4.5\,keV CRSF originally detected in the \chandra spectrum of M51 ULX-8 by \citet{brightmanMagneticFieldStrength2018}. We used the models reported by \citet{Middleton2019M51}, who carried out a more detailed analysis of the continuum (modelled with a soft {\sc diskbb} and a {\sc cutoffpl} for the high energy emission). We found that a 50\,ks \hexp observation detects the line at the $\sim$97\% (2.2$\sigma$) level while a 100~ks exposure detects the line in excess of 99.73\% (3$\sigma$) (to be compared with the 99.98\% or 3.8$\sigma$ detection by \chandra in 180\,ks). Fig.~\ref{fig:CRSF_lines} shows the spectrum as observed by \hexp.

Another unsolved problem \hexp will tackle is determining the population of particles producing the CSRF lines. In the case of electrons, further associated lines at the harmonic frequencies are expected, whereas protons are expected to produce lines only at the fundamental frequency (see \citealt{Staubert2019} and references therein). \citet{brightmanMagneticFieldStrength2018} were unable to detect any harmonic feature at 9\,keV, likely due to the low effective area of \chandra in this band. If we assume an additional line with the same properties (width and depth) as the fundamental but at twice the energy, the significance of the CRSF detection (now combined) would obviously increase (for 100\,ks, an additional $\sim \Delta \chi^2 = 26$ compared to the case where only the fundamental is present in the spectrum).
We found that an exposure of about 800\,ks using \hexp would allow us to detect the harmonic at the 99\% (2.6$\sigma$) level \textit{alone}. At this significance, combining both features will allow us to unambiguously identifying whether the CRSF is due to electrons or protons and thus accurately determine the magnetic field strength close to the NS.

\section{Conclusion}

As extragalactic sources, every ULX study -- population, spectroscopy, temporal analysis -- is affected by the sensitivity, angular resolution and instrumental background of X-ray telescopes.
All major discoveries in the field of ULXs have been driven by instruments combining focusing capabilities and effective area over the $\sim0.3-30$\,keV band, such as the combination of \xmm and \nustar.
However, many galaxies containing ULXs are still unresolved by these two instruments, e.g., M51 and NGC~253. In addition, the effectiveness of multi-instrument analysis is lowered by the difficulties of retrieving simultaneous observations.

\hexp's low- and high-energy telescopes ensure broadband simultaneous coverage of the X-ray spectrum and excellent angular resolution, significantly improved over both \xmm (for the LET) and \nustar (for the HETs).
Thanks to the lower background ensured by its small point-spread function, \hexp facilitates the identification of ULXs through their distinct curved spectra even at large distances.

The design of the \hexp mission would make it the ideal facility to shed light on a large number of questions regarding ULXs and super-Eddington accretion in general. In particular, broad-band and good energy resolution spectra would lead to a deep understanding of the balance between the mass inflow and outflow. \hexp will also enrich ULX population studies, for example by identifying the presence of NSs in ULXs through the detection of pulsations and CRSFs with significantly higher fidelity than existing missions.
As shown in Section~\ref{sec:pulsar}, \hexp surpasses the sensitivity of \nustar and \xmm in detecting pulsations from sources in crowded fields, especially ULXs. As a result, \hexp extends the range for detecting pulsars to distances 2 to 5 times farther than currently achievable, thereby probing a much larger (8--125 times) volume. This will permit access to a statistically significant population of bright extragalactic X-ray pulsars for the very first time.
Further demonstration of \hexp's capabilities in pulsar searches in crowded environments can be found in \citet{moriHighEnergyXray2023} (Galactic Center) and \citet{alfordHighEnergyXray2023} (Magnetars, CCOs).

In addition to the above, the L1 orbit of \hexp enables a continuous observation of a large fraction of the sky, without Earth occultation (an issue for \nustar, which is located in a low Earth orbit) or increased particle background from the Van Allen belts (an issue for \xmm and \chandra, with their highly elliptical orbits). This allows a significantly shorter ``clock time'' (actual time spent observing, as opposed to clean source exposure), while the broad bandpass of \hexp eliminates the issues due to differing observation windows when coordinating satellites in different orbits (\citealt{Middleton2017}).
This makes \hexp more efficient, enabling observations of more sources in a given amount of time, with much longer exposure times for observations constrained by clock time (e.g., coordinated observations, orbital measurements).

In summary, the \hexp mission concept holds tremendous promise for advancing our understanding of ULXs and the physics behind super-Eddington accretion. Its broad energy coverage, improved sensitivity, and exceptional spectral and timing capabilities establish it as a powerful tool for investigating ULXs and the fundamental physics at play.

\section*{Conflict of Interest Statement}

The authors declare that the research was conducted in the absence of any commercial or financial relationships that could be construed as a potential conflict of interest.

\section*{Author Contributions}

This Paper is the result of the work of the \hexp ULX working group, led by MBa and MJM. MBa, MJM, CP, AGL wrote the manuscript. DJW provided significant input throughout the design and writing process. MBa and BL wrote the NGC~253 simulation code based on SIXTE. MBa designed and produced the pulsar simulations. CP designed and produced the wind absorption/emission line simulations. AGL, DW, MBr and GV designed and produced the cyclotron line simulations. All authors have provided inputs during working group discussions and commented on the manuscript.

\section*{Funding}
MB was funded in part by PRIN TEC
INAF 2019 ``SpecTemPolar! -- Timing analysis in the era of high-throughput photon detectors''. TPR acknowledges funding from STFC as part of the consolidated grants ST/T000244/1 and ST/X001075/1. GV acknowledges support by Hellenic Foundation for Research and Innovation (H.F.R.I.) under the ``3rd Call for H.F.R.I. Research Projects to support Postdoctoral Researchers'' through the project ASTRAPE (Project ID 7802).  The work of DS was carried out at the Jet Propulsion Laboratory, California Institute of Technology, under a contract with NASA.

\section*{Acknowledgments}
The authors wish to thank the three reviewers for their comments, that helped improving the clarity and presentation of the manuscript.

\bibliographystyle{Frontiers-Harvard}
{\footnotesize
\bibliography{hexpulx}}

\end{document}